\documentclass[fleqn,usenatbib]{mnras}

\usepackage{newtxtext,newtxmath}

\usepackage[T1]{fontenc}

\DeclareRobustCommand{\VAN}[3]{#2}
\let\VANthebibliography\thebibliography
\def\thebibliography{\DeclareRobustCommand{\VAN}[3]{##3}\VANthebibliography}

\usepackage{graphicx}   
\usepackage{amsmath}	
\usepackage{CJKutf8}    
\usepackage{color,soul} 
\usepackage{verbatim}   
\usepackage{array}
\usepackage{pdflscape}  

\usepackage{tikz,xcolor,hyperref}

\newcommand{\mgii}{Mg\,\textsc{ii}}
\newcommand{\civ}{C\,\textsc{iv}}
\newcommand{\feii}{Fe\,\textsc{ii}}
\newcommand{\feiii}{Fe\,\textsc{iii}}

\newcommand{\lya}{Ly\textsc{$\alpha$}}
\newcommand{\hbeta}{H\textsc{$\beta$}}
\newcommand{\halpha}{H\textsc{$\alpha$}}

\defcitealias{Campitiello_2018}{C18}
\defcitealias{Nemmen_2010}{N10}
\defcitealias{Campitiello_2020}{C20}

\definecolor{lime}{HTML}{A6CE39}
\DeclareRobustCommand{\orcidicon}{%
    \begin{tikzpicture}
    \draw[lime, fill=lime] (0,0) 
    circle [radius=0.16] 
    node[white] {{\fontfamily{qag}\selectfont \tiny ID}};
    \draw[white, fill=white] (-0.0625,0.095) 
    circle [radius=0.007];
    \end{tikzpicture}
    \hspace{-2mm}
}

\newcommand{\orcidChrisO}{\href{https://orcid.org/0000-0003-0017-349X}{\orcidicon}}
\newcommand{\orcidChrisW}{\href{https://orcid.org/0000-0002-4569-016X}{\orcidicon}}
\newcommand{\orcidSamuel}{\href{https://orcid.org/0000-0001-9372-4611}{\orcidicon}}
\newcommand{\orcidFuyan}{\href{https://orcid.org/0000-0002-1620-0897}{\orcidicon}}

\title[Accretion disc fitting of SMSS J2157--3602]{Characterising SMSS J2157--3602, the most luminous known quasar, with accretion disc models}

\author[S. Lai et al.]{Samuel Lai \begin{CJK}{UTF8}{gbsn}(赖民希)\end{CJK},$^{1}$\orcidSamuel\thanks{E-mail: samuel.lai@anu.edu.au}
Christian Wolf,$^{1,2}$\orcidChrisW\,
Christopher A. Onken,$^{1,2}$\orcidChrisO\,
and Fuyan Bian \begin{CJK}{UTF8}{gbsn}(边福彦)\end{CJK}$^{3}$\orcidFuyan\,
\\
$^{1}$Research School of Astronomy and Astrophysics, Australian National University, Canberra, ACT 2611, Australia\\
$^{2}$Centre for Gravitational Astrophysics, Research Schools of Physics, and Astronomy and Astrophysics, Australian National University\\
$^{3}$European Southern Observatory, Alonso de C\'{o}rdova 3107, Casilla 19001, Vitacura, Santiago 19, Chile\\
}

\date{Accepted XXX. Received YYY; in original form ZZZ}

\pubyear{2022}

\begin{document}
\label{firstpage}
\pagerange{\pageref{firstpage}--\pageref{lastpage}}
\maketitle

\begin{abstract}
We develop an accretion disc (AD) fitting method, utilising thin and slim disc models and Bayesian inference with the Markov-Chain Monte-Carlo approach, testing it on the most luminous known quasar, SMSS J215728.21-360215.1, at redshift $z=4.692$. With a spectral energy distribution constructed from near-infrared spectra and broadband photometry, the AD models find a black hole mass of $\log(M_{\rm{AD}}/M_{\odot}) = 10.31^{+0.17}_{-0.14}$ with an anisotropy-corrected bolometric luminosity of $\log{(L_{\rm{bol}}/\rm{erg\,s^{-1}})} = 47.87 \pm 0.10$, and derive an Eddington ratio of $0.29^{+0.11}_{-0.10}$ as well as a radiative efficiency of $0.09^{+0.05}_{-0.03}$. Using the near-infrared spectra, we estimate the single-epoch virial black hole mass estimate to be $\log(M_{\rm{SE}}/M_{\odot}) = 10.33 \pm 0.08$, with a monochromatic luminosity at 3000\AA\ of $\log{(L(\rm{3000\text{\AA}})/\rm{erg\,s^{-1}})} = 47.66 \pm 0.01$. As an independent approach, AD fitting has the potential to complement the single-epoch virial mass method in obtaining stronger constraints on properties of massive quasar black holes across a wide range of redshifts.
\end{abstract}




\section{Introduction}
Active galactic nuclei (AGNs) with black holes weighing up to 10 billion solar masses have been discovered in the early Universe, less than a billion years after the Big Bang \citep{Ghisellini_2015, Wu_2015}. These luminous high-redshift quasars (QSOs) hosting massive black holes present significant challenges to theoretical models of supermassive black hole growth. Discovering and characterising luminous QSOs allows us to better understand the massive seeds and super-Eddington accretion scenarios necessary for the black holes to reach the observed masses \citep{Bromm_2003, Pacucci_2015, Amarantidis_2019, Zubovas_2021}. Furthermore, luminous QSOs at high-redshift are also useful for mapping our Universe's cosmic reionisation history \citep[e.g.][]{Banados_2018, Davies_2018}. 

As QSOs are powered by rapid accretion onto supermassive black holes (SMBHs), the thermal emission from an accretion disc typically peaks in the rest-frame ultraviolet (UV) and lends a degree of homogeneity to samples of QSO spectra. Complete samples of QSOs, however, have a tail towards redder UV-optical colours, usually attributed to dust obscuration \citep[e.g.][]{Webster_1995, Richards_2003, Glikman_2007, Klindt_2019, Fawcett_2022}. But also, accretion discs of more massive black holes are colder and thus their intrinsic spectra peak at longer wavelengths \citep[][]{Laor_Davis_2011}. This is because the innermost stable orbits of discs scale with the black hole mass $M_{\rm BH}$, while the thin accretion disc radius for a given temperature scales with $\sim M_{\rm BH}^{2/3}$ at fixed Eddington ratio. The most extreme SMBHs in QSOs might thus appear red in the rest-frame UV due to a large BH mass instead of dust. Disentangling the effects of BH masses and dust on QSO colours then relies on observing the disc over a sufficiently broad wavelength range.

A number of different methods have been employed in recent years to estimate black hole masses ($M_{\rm{BH}}$) including: reverberation mapping \citep[e.g.][]{Blandford_1982_RM, Peterson_1993, Netzer_1997, Wandel_1999, Kaspi_2000, Peterson_2004}, velocity-delay maps \citep{Grier_2013, DeRosa_2018}, gravitational micro-lensing \citep[e.g.][]{Irwin_1989, Sluse_2011, Guerras_2013}, polarization of broad emission-lines \citep[e.g.][]{Savic_2018, Savic_2021, Capetti_2021, Popovic_2022}, single-epoch virial estimates \citep[e.g.][]{Vestergaard_2002, Mclure_2002, Mclure_2004, Greene_2005, Vestergaard_2006}, accretion disc fitting \citep[e.g.][]{Malkan_1983, Wandel_1988, Sun_1989, Laor_1990, Rokaki_1992, Tripp_1994, Calderone_2013, Capellupo_2015, Campitiello_2018, Mejia_Restrepo_2018, Cheng_2019}, dynamical estimates \citep[e.g.][]{Davies_2006_dynamical, onken_2007_dynamical, Hicks_2008, Greene_2010, Raimundo_2019}, and gravitational redshift of the \feiii\, line \citep{Mediavilla_2018, Mediavilla_2019}. These methods are widely applied based on the features of the system being studied and the availability of data on various observables. In particular, the single-epoch virial mass estimate is frequently used on large samples across a wide range of redshifts \citep[e.g.][]{Shen_2011}. It is also the most common method applied for QSOs found at the highest redshifts \citep[e.g.][]{Wu_2015, Mazzucchelli_2017, Reed_2019, Wang_2021} due to the relative ease of obtaining single-epoch spectra compared to spatially or temporally resolved observations.

In this study, we characterise SMSS J215728.21$-$360215.1 (hearafter J2157--3602), the most luminous known quasar, with a spectroscopically measured redshift of $z = 4.692$ \citep{Wolf_2018, onken_2020_J2157}. Using both survey photometry and spectra, we construct a spectral energy distribution (SED) from the rest-frame infrared to ultraviolet. We describe further developments to the accretion disc fitting technique by Bayesian inference of black hole mass with the Markov-Chain Monte-Carlo sampling approach and compare our measurements against the single-epoch virial method. Constraining the peak of the accretion disc emission proves to be effective at measuring the black hole mass \citep{Campitiello_2018}. For the peak accretion disc emission to be located at wavelengths longer than \lya, the black hole needs to be sufficiently high mass. Our results show that AD fitting is a viable independent method for characterising the highest mass QSO black holes, which can be applied even in the absence of spectra, making it suitable for large samples. This creates an opportunity to obtain more stringent constraints on black hole properties and their growth rates, particularly for high-redshift QSOs where other techniques are impractical.

The content of this paper is organised as follows: in Section \ref{sec:SED}, we describe the photometric and spectroscopic data obtained for J2157--3602. In Section \ref{sec:characterisation_methods}, we present the two complementary methods we use in this study to characterise black hole mass. In Section \ref{sec:results_discussion}, we discuss measurements of the black hole mass and bolometric luminosity, in addition to estimates of the Eddington ratio and radiative efficiency. We conclude with Section \ref{sec:conclusion}. Throughout the paper, we adopt a flat $\Lambda$CDM cosmology with H$_{0} = 70$ km s$^{-1}$ Mpc$^{-1}$ and $\left(\Omega_{\rm m}, \Omega_{\Lambda}\right) = \left(0.3, 0.7\right)$. All referenced emission-line wavelengths are measured in vacuum.

\section{Spectral Energy Distribution}
\label{sec:SED}
We use spectroscopic observations and publicly available survey photometry to construct the SED of J2157--3602. 

\subsection{Photometric Data}
We crossmatch J2157--3602, positioned at RA = 329.36762$^{\circ}$ and Dec = -36.03756$^{\circ}$ (J2000), with AllWISE \citep{WISE, ALLWISE}, the VISTA Hemisphere Survey \citep[VHS;][]{VHS} DR6, Two Micron All-Sky Survey \citep[2MASS;][]{2MASS}, SkyMapper Southern Survey \citep[SMSS;][]{SMSS_2019} DR3, and NOIRLab Source Catalog \citep[NSC;][]{NSC_DR2} DR2 to collect photometry from the infrared to optical passbands. The W1 and W2 magnitudes from CatWISE2020 \citep{Marocco_2021_catwise2020} are not appreciably different from that of AllWISE. Properties of J2157--3602, including its photometry, can be found in the discovery paper \citep{Wolf_2018}. We also obtain the transmission profile of all broadband filters using the SVO Filter Profile Service \citep{SVO_Filter_Profile_Service}.

\subsection{Spectroscopic Data}

Comprehensive details of the spectroscopic observation and data description are presented in \citet{onken_2020_J2157}. Briefly, observations were obtained from two medium resolution and wide-band spectrographs: the Near-Infrared Echellette Spectrometer (NIRES) instrument \citep{Wilson_2004_NIRES} at Keck Observatory and the X-shooter instrument \citep{Vernet_2011_Xshooter} at the Very Large Telescope. The full observed wavelength coverage is from 3000 \AA\ to nearly 2.5 $\mu$m, although little of the source flux is transmitted shortward of \lya\ at an observed wavelength of $\sim6920$~\AA. The data are reduced using \texttt{PypeIt} \citep{Pypeit_2020} and a stacked spectrum is created by scaling the NIRES spectrum to the X-shooter data in the overlapping wavelength region. The absolute flux calibration of the combined spectrum is based on the VHS DR6 J-band photometry. From the \mgii\ line, the systemic redshift is measured to be $z = 4.692$ and the median signal-to-noise (SNR) per $\sim$50 km s$^{-1}$ velocity dispersion bin measured between rest-frame 2700$-$2900\AA\ is nearly 200.

\subsection{Galactic Extinction} \label{sec:Data_Extinction}
For Galactic extinction, we use $R_{\rm{v}} = 3.1$ and the Schlegel, Finkbeiner \& Davis \citep[SFD;][]{Schlegel_1998} extinction map to apply a correction in the observed frame. We also utilise a 14\% re-calibration factor $E(B-V) = 0.86 \times E(B-V)_{\rm{SFD}}$ \citep{Schlafly_2011}, which is informed by the Sloan Digital Sky Survey \citep[SDSS;][]{York_2000_SDSS} data and an analysis of the blue tip of the stellar locus \citep{Schlafly_2010}. We find the Galactic extinction to be small for J2157--3602, confirmed by $E(B-V) = 0.013$, and due to its extreme luminosity, we assume no host galaxy extinction. However, we briefly discuss the effect of host galaxy extinction on the $M_{\rm{BH}}$ measurement in Section \ref{sec:dust_ext}.

\section{Black Hole Characterisation} \label{sec:characterisation_methods}
We discuss two complementary methods for estimating the black hole mass of J2157--3602 from photometric and spectroscopic data.

\subsection{Single-Epoch Virial Mass}
We begin by describing the single-epoch (SE) virial mass method, in order to later examine use of the virial mass estimate to inform the Bayesian priors of the accretion disc fitting. The SE technique is routinely applied to QSO spectra \citep[e.g.][]{Vestergaard_2002, Mclure_2002, Mclure_2004, Greene_2005, Vestergaard_2006}. The dynamics of the line-emitting gas is assumed to be virialised with the gravitational potential of the black hole. The velocity-broadened emission-line profile measures the gas velocity, and the continuum luminosity is used to infer the radius of the broad-line region (BLR) through the radius-luminosity (R-L) relation, which is empirically derived from reverberation mapping experiments \citep[e.g.][]{Kaspi_2000, Kaspi_2005, Bentz_2006, Bentz_2013}. While most reverberation mapping experiments calibrate the R-L relation using the \hbeta\ line, the \mgii\ line profile is found to correlate with \hbeta\ and can be used as its substitute \citep[e.g.][]{Salviander_2007, Shen_2008, Wang_2009, Shen_2012}. This is convenient for the J2157--3602 spectrum where the \hbeta\ line is redshifted out of the X-shooter NIR coverage into wavelengths that are more difficult to observe from the ground. Single-epoch virial mass estimates take on the following form, 
\begin{equation}
   \left(\frac{M_{\rm{SE}}}{M_{\odot}}\right) = 10^{\rm{a}} \left[\frac{\lambda L_{\lambda}}{10^{44} \,\rm{erg\, s^{-1}}}\right]^{b} \left[\frac{\rm{FWHM_{\rm{line}}}}{1000 \,\rm{km\, s^{-1}}}\right]^{c} \,,
   \label{eq:mgii_virial}
\end{equation}
where $\lambda L_{\lambda}$ is the monochromatic luminosity at a particular wavelength or an emission line luminosity and FWHM is the full-width half-maximum of a broad emission line. In this study, we use the \citet{Shen_2011} \mgii\ calibration with exponents (a, b, c) calibrated to the values (6.74, 0.62, 2.0), based on a high-luminosity local AGN subsample. The differences between sets of calibrations, such as those from \citet{Vestergaard_2009}, are approximately 0.1 dex, but the overall statistical uncertainties of the SE virial mass estimate can be up to 0.5 dex, due to QSO variability and the propagated uncertainties from reverberation mapping masses \citep[e.g.][]{Krolik_2001, Woo_2010, Steinhardt_2010, Shen_2013, Kozlowski_2017, DallaBonta_2020}. In this study, we present the SE virial mass estimates with their measurement uncertainties, but acknowledge that the 0.5 dex statistical uncertainty is often the dominant error.

\subsubsection{Spectral fitting}
For the SE mass estimate, we fit the stacked spectrum from Keck/NIRES and VLT/X-shooter. One of the primary difficulties in obtaining the pure velocity-broadened profile of \mgii\ is the treatment of the broad \feii\ emission, which forms a pseudo-continuum in QSO rest-frame UV and optical spectra. Following after similar studies of QSO spectra \citep[e.g.][]{Wang_2009}, we model the underlying continuum with three components: a power-law, Balmer continuum, and blended \feii\ flux. Our power-law is normalised to 3000\AA\ while the Balmer continuum is modeled as a Planck blackbody with a uniform electron temperature $T_{\rm{e}}$ attenuated by an optical depth $\tau_{\lambda}$ \citep[e.g.][]{Grandi_1982, Dietrich_2002, Wang_2009, Kovacevic_2014}. Both parameters, along with the overall normalisation, are free parameters of the continuum fit. The Balmer continuum is often not well constrained independently of the power-law and \feii\ continuum, so we do not suggest a physical interpretation of the temperature and optical depth. 

The \mgii\ line profile is sensitive to the underlying \feii\ emission features. Adopting only one \feii\ model can induce a bias in the resulting FWHM \citep[e.g.][]{Schindler_2020}, so we consider a variety of empirical and semi-empirical \feii\ templates. Our \citet[][VW01]{Vestergaard_2001} template is spliced with \citet{Salviander_2007}, which extrapolates under the \mgii\ line from the rest-frame 2200$-$3090\AA; it is the same version of VW01 used in other spectral fitting codes such as \texttt{PyQSOFit} \citep{Guo_2018}. We also use the \citet[][T06]{Tsuzuki_2006}, \citet[][BV08]{Bruhweiler_Verner_2008}, and \citet[][M16]{Mejia-Restrepo_2016} templates. 

The full pseudo-continuum is uniquely defined by eight free parameters. All components of the continuum are simultaneously fit to selected windows in close proximity to the \mgii\ line: 2200$-$2740\AA, 2840$-$3300\AA, and 3500$-$3650\AA\ in the rest-frame. The emission-line is fit to the continuum-subtracted spectrum between the rest-frame wavelengths 2710$-$2930\AA, which overlaps with the continuum windows in order to force the \mgii\ flux contribution to converge to zero at the wings. Similar results (within 0.5$\sigma$) are obtained without the overlap. The broad line is fit with a maximum of three Gaussian components, each with an independent wavelength shift of up to $\pm$30\AA. Although the decomposition of the \mgii\ line is not necessarily unique, we obtain the FWHM from the total line profile, which is less sensitive to the particulars of the decomposition.

\subsection{Accretion Disc Fitting} \label{sec:ADfitting}
The QSO accretion disc emission can be modeled by a superposition of blackbodies with a wide range of effective temperatures, where the spatial temperature profile is derived from the disc emissivity under classical accretion disc theory \citep{SS73, Novikov_Thorne_1973}. Under the standard framework of assuming large optical thickness, gas radiates isotropically as a blackbody with a local effective temperature, and flux is integrated across the whole disc surface down to the innermost edge. For a set of black hole properties, its mass and spin, alongside an observed inclination angle and accretion rate or disc luminosity, one can produce predicted SEDs of the observed thermal emission by ray-tracing null geodesics from an observer placed at infinity to the vicinity of the black hole. The steady-state accretion disc structure is solved by semi-analytical or numerical models. The accretion disc fitting method (hereafter AD fitting) can be used to recover black hole properties from observed spectra \citep[e.g.][]{Kawaguchi_2004}. Specifically, we use AD fitting to estimate the black hole mass, $M_{\rm{AD}}$, where the uncertainty of the measurement originates from a space of degenerate solutions due to the unknown black hole spin and orientation. 

\subsubsection{Spectral hardening}
Realistic accretion disc emission is more complex than a superposition of blackbodies at various temperatures, motivating the addition of a colour correction (hardening) factor, $f_{\rm{col}} \geq 1$, to capture the combined effects of Compton scattering, absorption opacity, optical depth, and plasma density structure \citep[e.g.][]{Shimura_1993, Ebisawa_1993, Shimura_1995}. In practice, $f_{\rm{col}}$ approximates the departure of a more realistic disc from a multitemperature blackbody model. The black hole mass, $M_{\rm{AD}}$, is degenerate with the colour correction, following the relation $M_{\rm{AD}} \propto f_{\rm{col}}^2$ \citep{Ebisawa_1993, Shimura_1995} for a Keplerian thin disc with all other intrinsic properties being equal \citep[][Eq. 16-17]{Li_2005}. Thus, any black hole mass measurements from AD fitting will depend sensitively on the hardening factor.

The value of $f_{\rm{col}}$ is most sensitive to the accretion disc's maximum effective temperature $T_{\rm{max}}$ \citep{Davis_2005, Done_2012, Davis_El-Abd_2019}, where $f_{\rm{col}}$ converges toward unity for $T_{\rm{max}} \leq 3 \times 10^4 \,\rm{K}$ \citep{Davis_2011}. At sufficiently low temperatures, helium is largely in a neutral state and the enhanced neutral hydrogen fraction increases absorption opacity, supporting the complete thermalisation of the accretion disc while minimising scattering and Comptonisation effects from free electrons. The maximum effective accretion disc temperature is proportional to accretion rate as $\dot{M}^{1/4}$ and to black hole mass as $M_{\rm{BH}}^{-1/2}$ \citep{Laor_Davis_2011}. This suggests that the $f_{\rm{col}}$ correction is high for low-mass black holes accreting at super-Eddington rates, where electron scattering is dominant. In contrast, for supermassive black holes, such as J2157--3602, where $\log(M_{\rm{BH}}/M_{\odot}) > 8$ and Comptonisation has little effect \citep{Hubeny_2001}, the totality of the accretion disc radiation is likely dominated by thermal emission.

In contrast to the typical value of $f_{\rm{col}} = 1.7$, which was developed for X-ray binaries (XRBs) \citep{Shimura_1995}, similar studies of AGN with supermassive black holes adopt $f_{\rm{col}} = 1$ \citep{Vasudevan_2007, Vasudevan_2009, Done_2012, Calderone_2013, Campitiello_2018, Mejia_Restrepo_2018, Cheng_2019}. However, we caution that some models predict an increased hardening factor with black hole mass \citep[e.g.][]{Davis_El-Abd_2019}. Other approaches and theoretical integrated spectra of AGN accretion discs around supermassive BHs predict that the hardening factor should be close to unity \citep{Hubeny_2000, Hubeny_2001, Czerny_2011}. Empirically, if the hardening factor was significant, AD fitting studies that assume a value of unity should find their measurements of $M_{\rm{BH}}$ to be systematically underestimated compared to independent measurements, but significant overestimation is sometimes observed instead \citep[e.g.][]{Calderone_2013, Campitiello_2020}. In this study, we assume $f_{\rm{col}} = 1$ to remain consistent with similar AD fitting approaches developed for AGN with supermassive BHs.

We consider both geometrically thin \citep[\texttt{kerrbb};][]{Li_2005} and slim \citep[\texttt{slimbh};][]{Sadowski_2011, Straub_2011} accretion disc emission models, where the pre-calculated spectral tables are accessed through \texttt{XSPEC} 12.12.0 \citep{XSPEC}, packaged as part of \texttt{Sherpa} \citep{Sherpa}, the modeling and fitting suite of Chandra Interactive Analysis of Observations (CIAO) v4.14. Both models are ray-traced numerical solutions to steady-state accretion disc models with general relativistic effects and they have been found to be consistent with synthetic spectra created from numerical general relativistic radiative magnetohydrodynamic simulations of puffy discs \citep{Wielgus_2022}. To tackle the inherent degeneracies in the problem, we use Markov-Chain Monte-Carlo techniques to infer the black hole mass from its posterior probability distribution.

\subsubsection{Thin disc}
The \texttt{kerrbb} model is an optically thick and geometrically thin numerical Keplerian accretion disc model originally designed for black hole X-ray binary spectra \citep{Li_2005}. Ray-tracing is used to calculate the spectrum, under the assumption that emission from every point in the accretion disc is locally blackbody-like. A limitation is that the \texttt{kerrbb} model assumes a perfectly flat disc with no vertical thickness, flaring, or warping \citep{Li_2005}. Nevertheless, \citet{Campitiello_2018}, hereafter \citetalias{Campitiello_2018}, used \texttt{kerrbb} to develop analytic expressions approximating the black hole mass $M_{\rm{BH}}$ and accretion rate $\dot{M}$ from fitting the QSO SED, finding that $M_{\rm{AD}}$ can be constrained by the peak of the accretion disc spectrum, $\nu_{\rm{p}}$, and the peak luminosity, $\nu_{\rm{p}}L_{\nu_{\rm{p}}}$. However, even for a fixed inclination $\theta_{\rm{inc}}$, the peak frequency and luminosity are degenerate with $M$, $\dot{M}$, and $a$, such that any accretion disc SED can be reproduced by a carefully selected family of solutions. The parameter space covered by the degenerate solutions can be minimised by making reasonable assumptions about the spin and inclination \citepalias{Campitiello_2018}. Despite the degeneracies involved, we discuss how we derive the maximum likelihood black hole mass in Section \ref{sec:mcmc}. 

Comparisons of reverberation mapping and SE virial mass estimates showed that the analytical approximations of \citetalias{Campitiello_2018} can measure $M_{\rm{AD}}$ to a precision of 0.45 dex \citep[][hereafter \citetalias{Campitiello_2020}]{Campitiello_2020}. They further showed that the \texttt{kerrbb}-based mass estimates are systematically larger (up to 0.4 dex depending on black hole mass and spin) than masses estimated through SE methods for their sample, which spans $\log{(\rm{M_{BH}}/\rm{M_\odot})} = 7.5-9.5$.  

In our application of \texttt{kerrbb}, we toggle on the effects of self-irradiation, whereby radiation can be gravitationally deflected, illuminating another part of the accretion disc, and include limb-darkening effects. We also set the torque at the inner boundary of the accretion disc to zero, following the standard theory of accretion discs. The effect of a nonzero torque enhances the disc power by boosting returning radiation, particularly at higher energies. However, nearly indistinguishable spectra can be created by adjusting the effective mass accretion rate even for very high torques \citep{Li_2005}. We show in Section \ref{sec:BHmassAD} an alternative $M_{\rm{AD}}$ estimate if the disc power from torque is comparable to the disc emission.

\subsubsection{Slim disc}
The thin disc approximation breaks down for Eddington ratios $\geq 0.3$ \citep[e.g.][]{Laor_1989}, motivating the relaxation of the thin disc assumption. The \texttt{XSPEC} spectral model, \texttt{slimbh}, is a fully relativistic, optically thick slim disc accretion model with numerical ray-tracing originally designed to fit the X-ray continuum of black hole X-ray binaries \citep{Sadowski_2011, Straub_2011}. In contrast to standard thin disc models, \texttt{slimbh} relaxes the imposed Keplerian angular momentum condition and accounts for advective cooling \citep{Sadowski_2011}. \citet{Campitiello_2019} derived analogous analytical approximations showing that, as with the thin disc model, the black hole mass can be constrained by the peak flux and wavelength of the accretion disc SED. However, in the \texttt{slimbh} model, the mass accretion rate $\dot{M}$ is replaced with the disc luminosity $L_{\rm{disc}}$ parameterised in terms of the Eddington ratio $\lambda_{\rm{Edd}}$, which introduces a black hole mass dependence. Additionally, the vertical structure is only evaluated up to the Eddington limit. In the low Eddington regime, \texttt{slimbh} spectra converges towards the geometrically thin \texttt{kerrbb} spectra, making it a more general model, except that zero torque is enforced at the inner boundary of the accretion disc. Another constraint is that \texttt{slimbh} is restricted to positive black hole spins only, whereas \texttt{kerrbb} is also calculated for negative spins. 

\begin{figure*}
	\includegraphics[width=1.0\textwidth]{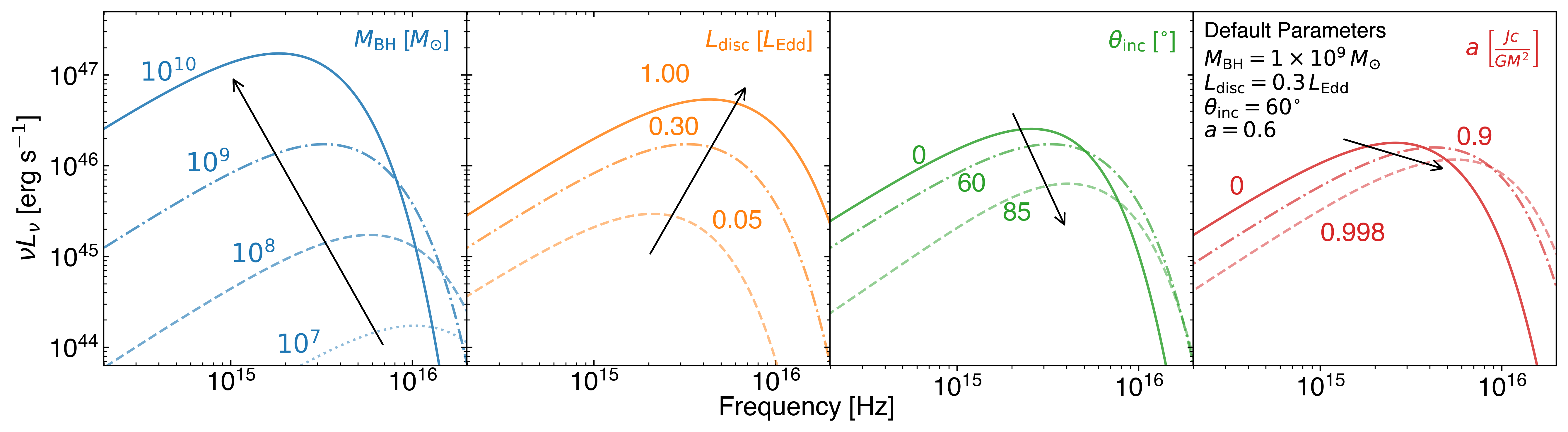}
    \caption{Accretion disc spectra created from \texttt{slimbh} slim disc models, with each panel varying one of the four parameters: (from left to right) black hole mass, disc luminosity, inclination, and spin. The default parameters are listed in black within the rightmost panel and all SEDs are labelled with the modified value corresponding to the parameter on the top-right of each panel. The black arrows indicate the direction that the SEDs evolve towards as the parameter increases for this test case.}
    \label{fig:Sample-SEDs}
\end{figure*}

In our application of \texttt{slimbh}, we toggle on the effects of limb darkening and the heightened surface profile, such that ray-tracing can be performed from the disc photosphere rather than from the equatorial plane. We assume a constant $f_{\rm{col}} = 1$ and do not utilise \texttt{BHSPEC} disc atmosphere models \citep{Davis_2005}. We also set the $\alpha$-viscosity parameter to 0.01. In Section \ref{sec:BHmassAD}, we discuss how the $M_{\rm{AD}}$ estimate is not sensitive to the $\alpha$-viscosity parameter. We present sample slim disc spectra created using \texttt{slimbh} for various black hole masses, disc luminosities, inclinations, and spins in Figure \ref{fig:Sample-SEDs}.

\subsubsection{Accretion disc SED} \label{sec:AD_SED}

We consider three datasets for fitting the AD models: synthetic photometry created from the stacked spectra, survey photometry,  and modified photometry with QSO template subtraction. In all cases, we apply the Galactic extinction correction from Section \ref{sec:Data_Extinction}. We fit accretion disc emission in a region bounded by rest-frame 1 $\mu$m and 1600\AA, due to the increasing hot dust contribution at the red end and the intergalactic medium \lya\ absorption at the blue end. Because of the wide wavelength coverage of each of the photometric data and broad \civ\ visible in the spectrum, characteristic of a high-ionisation broad absorption line (HiBAL) QSO, we leave a sizable buffer between the blue threshold and the \lya\ forest. Very broad photometry, such as those from \textit{Gaia}, are ill-suited to the accretion disc fitting approach because they can hide many details of the underlying spectral shape that are crucial to obtaining a good fit. 
\begin{enumerate}
    \item Synthetic Photometry (SynPhot)
    
    \noindent The presence of broad emission lines, the Balmer continuum, and \feii\ flux implies that a significant fraction of the structure observable in spectra is not required for fitting the underlying continuum emission. Instead of fitting the spectrum directly, we create synthetic photometric data points with a flat transmission profile for contiguous sections of accretion disc emission that are free of emission-line contribution. The emission-line-free windows are created by masking all broad lines and identifying regions of the Gaussian-broadened \citet[][BV08]{Bruhweiler_Verner_2008} \feii\ template that fall below a flux threshold, which we set to 70\% of the median template flux. Both the Gaussian broadening of the \feii\ template and the width of the broad line mask are scaled to a velocity dispersion value, which we set to 4500 km s$^{-1}$. The size of the mask is three times the velocity dispersion. We found that our results are not sensitive to reasonable values chosen for the \feii\ template flux threshold and width of the Gaussian broadening kernel. This procedure creates synthetic photometry from the stacked spectrum, which we use in place of the photometry from the same wavelength region. 

    For the red end of the SED outside of the wavelength coverage of the stacked spectrum, we scale the Selsing X-shooter template \citep{Selsing_2016} to match the flux of J2157--3602 photometry, where each photometric bandpass is treated independently. The cutouts of the Selsing template, each with the width of one bandpass, are used to create synthetic photometry using the same method as above. In this way, we replace the W1 and W2 photometric points, avoiding flux contribution from bright emission lines, such as \halpha. Because each photometric point is treated independently, differences in the global continuum slope between our target and the Selsing template are minimised. We test this by creating variations of the synthetic photometry for deviations in the Selsing template continuum slope, parameterised by an $F_{\lambda}$ power-law slope of $\alpha = -1.70$ \citep{Selsing_2016}. We test variations of $1\,\sigma_{\alpha}$, where we take $\sigma_{\alpha} = 0.6$ from the \citet{Rakshit_2020} catalogue, and find median flux variations of 5$-$7\% in the synthetic photometry We find the flux variations to have no significant effect on our measurements as described in Section \ref{sec:BHmassAD}.

    \item Photometry-Only
    
    \noindent For comparison to the synthetic photometry, we present results obtained from fitting only the survey photometry in Section \ref{sec:BHmassAD}.

    \item Template-Corrected Photometry
    
     \noindent In a further modification to the photometry-only models, we apply the same approach as for the long-wavelength data of the SynPhot model and scale the Selsing template to the observed photometry and use its continuum model to correct for flux contribution from emission lines. 
\end{enumerate}

These two alternative datasets to the synthetic photometry demonstrate how the black hole mass estimate from AD fitting is affected if spectroscopic data is unavailable.

\subsubsection{Markov-Chain Monte-Carlo} \label{sec:mcmc}
The free parameters in AD fitting are the black hole mass $M_{\rm{BH}}$, luminosity, spin $a$, and observer's inclination $\theta_{\rm{inc}}$ with respect to the symmetry axis of the accretion disc (equivalent to the spin axis), where the luminosity is parameterised in terms of the mass accretion rate $\dot{M}$ for \texttt{kerrbb} and the Eddington ratio for \texttt{slimbh}. The inclination angle corresponds to $\theta_{\rm{inc}} = 0^{\circ}$ when viewed face-on and $\theta_{\rm{inc}} = 90^{\circ}$ when viewed edge-on. These parameters are inescapably degenerate in the absence of independent constraints. We utilise Bayesian Markov-Chain Monte-Carlo (MCMC) methods, as implemented in the Python \texttt{emcee} module \citep{emcee}, to probe the multi-dimensional parameter space and obtain a posterior distribution of $M_{\rm{BH}}$. 

We adopt an uninformative, flat prior probability distribution for all free parameters, except for the black hole mass. There, we investigate three probability distributions: a flat prior, a SE virial mass prior, and a black hole mass function prior. The SE virial mass prior is a Gaussian distribution centred around the SE mass estimate with a width set to a statistical uncertainty of 0.5 dex. The black hole mass function prior is based on a double power-law parameterisation of the QSO luminosity function \citep{Onken_2022_QLF}, which is transformed from 145nm absolute magnitude to black hole mass through the \citet{Runnoe_2012} bolometric correction and an assumed fixed mean Eddington ratio. For mean Eddington ratios $> 0.1$, such as 0.25 \citep{Kollmeier_2006} or 0.6 \citep{Trakhtenbrot_2011}, the slope of the mass function on the bright-end, which determines the relative likelihood between black hole masses, is constant across the range of black hole masses being considered and the exact Eddington ratio assumed is inconsequential. We use double power-law parameters derived from \citet{Onken_2022_QLF} with faint-end constraints from \citet{Niida_2020} and the redshift evolution of the normalisation from \citet{Fan_2001} for the sample with a median redshift of $z = 4.83$. The location of the knee in the mass function for fixed mean Eddington ratios of 0.25 and 0.6 are $\log(M_{\rm{BH}}/M_{\odot}) = 9.28$ and $8.90$, respectively. For bright QSOs similar to J2157--3602, the slope of the mass function is a conservative prior which disfavours spuriously high black hole masses. 

In Section \ref{sec:BHmassAD}, we discuss the effect of the prior on our black hole mass measurement. The black hole spin parameter is allowed to explore the space of all positive spins, 0 < $a$ < 0.998, and we exclude edge-on cases for the inclination, requiring $\theta_{\rm{inc}} < 65^{\circ}$, which is our assumed opening angle of the obscuring torus. We consider edge-on orientations to be intrinsically less likely given the extreme luminosity of J2157--3602. We also check the convergence of our model using \texttt{ChainConsumer} \citep{Chainconsumer} and confirm that our $M_{\rm{BH}}$ posterior is independent of the initial parameters. The final $M_{\rm{AD}}$ estimates are taken from the maximum likelihood of the posterior distribution, with uncertainties determined by the 68\% iso-likelihood line. If a parameter could not be constrained this way such as when the distribution is strongly skewed towards one end of the allowed parameter space, then the median of the distribution is used instead, with uncertainties determined by the 16$^{\rm{th}}$ and 84$^{\rm{th}}$ percentiles. 

\section{Results and Discussion} \label{sec:results_discussion}

In this section, we present the results of both methods for determining $M_{\rm{BH}}$ from the combined photometric and spectroscopic data. We compare these methods against each other and to previous results obtained for the black hole of J2157--3602 \citep{onken_2020_J2157}. We discuss bolometric corrections to the observed luminosity at 3000\AA\ with anisotropy correction and use these quantities to further derive the QSO Eddington ratio and radiative efficiency. Finally, we discuss the potential of the AD fitting method.

\subsection{Black Hole Mass from SE Virial Method} \label{sec:BHmassSE}
Previous estimates of the black hole in J2157--3602 based on a similar analysis of the \mgii\ line found $\log(M_{\rm{SE}}/M_{\odot})_{\rm{\mgii}} = 10.53 \pm 0.08$ \citep{onken_2020_J2157}, using the \citet{Shen_2011} calibration. This value, however, depends on the FWHM measured above a continuum defined with the VW01 template, which can result in systematically overestimated FWHMs compared to other \feii\ templates \citep{Schindler_2020}.

\begin{figure}
	\includegraphics[width=1.0\columnwidth]{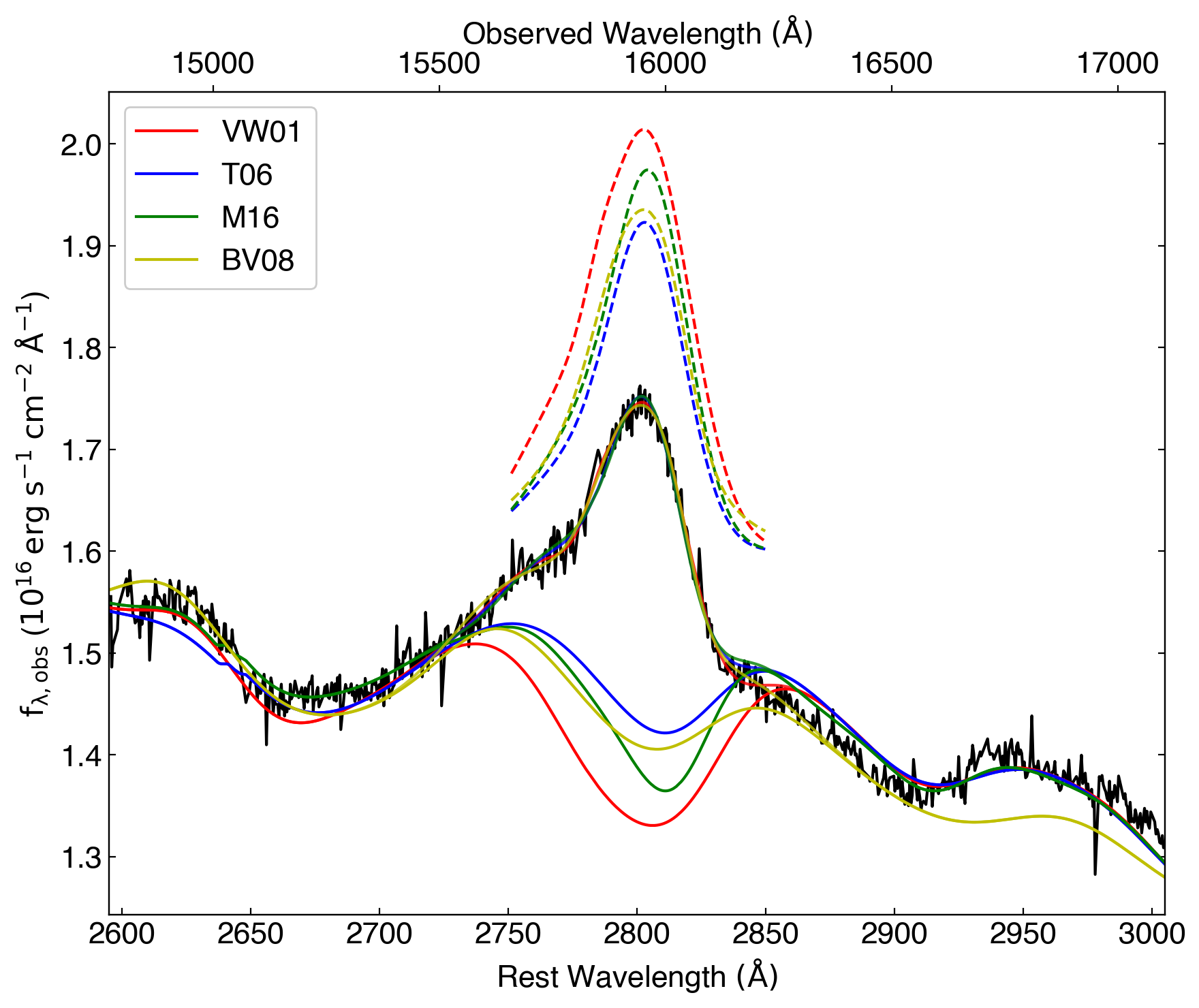}
    \caption{Fits of the \mgii\ emission feature of J2157--3602 with the combined pseudo-continuum from the power-law, Balmer, and \feii\ components. The fits differ from one another by the applied \feii\ template (see legend). The resulting continuum for each template is plotted in solid lines and the fitted line profile in dashed lines, while the data is shown in black. The BV08 template is not well-matched to features on the red end of the \mgii\ line. Individual emission line fits are offset to $1.6 \times 10^{-16}$ erg s$^{-1}$ cm$^{-2}$ \AA$^{-1}$ and the FWHM between fits with different \feii\ models range from 4200 to 5150 km s$^{-1}$. The \mgii\ FWHM mean and standard deviation are 4550 $\pm$ 400 km s$^{-1}$.}
    \label{fig:Fe_templates}
\end{figure}

Figure \ref{fig:Fe_templates} presents fits to the \mgii\ emission feature of J2157--3602 with various \feii\ templates. Each template is broadened by a Gaussian kernel with a dispersion between 3300--5100 km s$^{-1}$ depending on the template and the \mgii\ FWHM is measured in the range 4200--5150 km s$^{-1}$.

We consider two sources of uncertainty in the measurement: the statistical uncertainty from the choice of \feii\ template and the measurement uncertainty. We estimate the statistical uncertainty by independently fitting the emission feature with each of the four \feii\ templates and measuring variance in the resulting FWHMs. To minimise bias towards any particular template, the mean from the four fits is used as the final estimate of the FWHM. We also consider an additional measurement uncertainty which is estimated from fitting 100 synthetic spectra, where the flux at each wavelength bin is resampled according to a Gaussian distribution with a standard deviation equivalent to the flux uncertainty. The variance in the resulting fits is added to the final uncertainty of the FWHM. Because of the high SNR of the spectra, the statistical uncertainty from the \feii\ templates is over 5$\times$ the measurement uncertainty, but the final error in $M_{\rm{SE}}$ remains significantly below the overall statistical error of the virial mass estimator. 

From the Galactic extinction-corrected and stacked spectrum, we remeasure the FWHM to be 4550 $\pm$ 400 km s$^{-1}$ with fits to the \mgii\ feature, relative to 5720 $\pm$ 570 km s$^{-1}$ from \citet{onken_2020_J2157}. The monochromatic luminosity at 3000\AA\ measured from the power-law continuum is $\log{(L(\rm{3000\text{\AA}})/\rm{erg\,s^{-1}})} = 47.66 \pm 0.01$. Therefore, the adjusted black hole mass calculated from the SE virial mass estimate is $\log(M_{\rm{SE}}/M_{\odot})_{\rm{\mgii}} = 10.33 \pm 0.08$, which is 0.2 dex less massive than the estimate by \citet{onken_2020_J2157}. 

\subsection{Black Hole Mass from AD Fitting Method} \label{sec:BHmassAD}

We present the results from the AD fitting method in Table \ref{tab:ADmasses} for three choices of the prior probability distribution: a flat uninformative prior, a distribution based on the SE virial mass estimate, and another based on the QSO mass function. We also present fits to three datasets: the synthetic photometry (SynPhot), template-subtracted photometry (Template), and photometry-only (Photometry). Descriptions of these three datasets can be found in Section \ref{sec:AD_SED}. 

\begin{figure*}
	\includegraphics[width=0.90\textwidth]{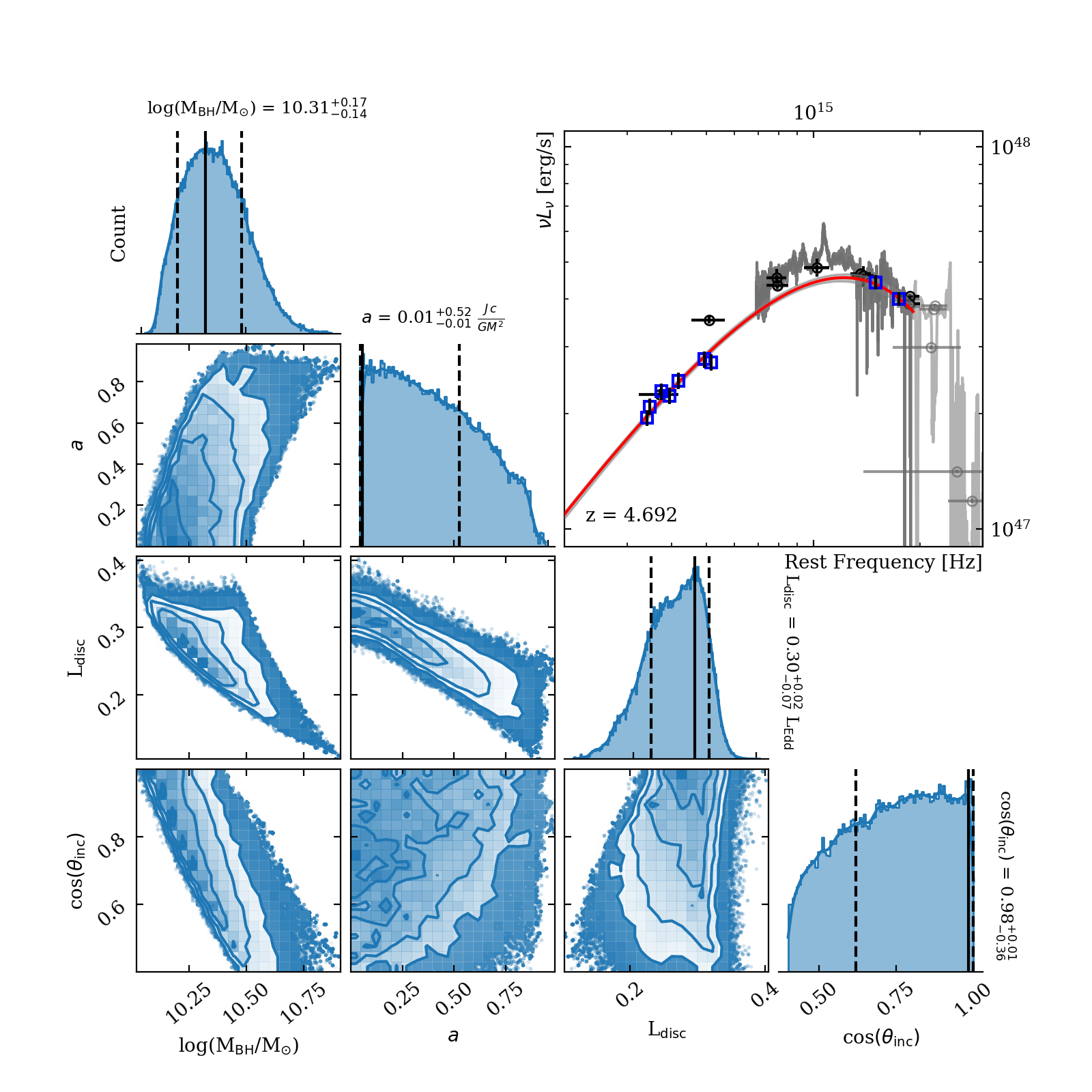}
    \caption{Corner plot for the accretion disc fit to J2157--3602 synthetic photometry using the \texttt{slimbh} model. The frequency is presented in the rest-frame. The SE virial mass estimate is used as a prior probability distribution on the black hole mass. The final estimate of each parameter is determined by the maximum likelihood of the posterior distribution, with uncertainties determined by the 68\% iso-likelihood line. The top-right panel shows the stacked spectrum in dark grey, the synthetic photometry in blue square points, photometry in black, and masked data in light grey. The red model is the highest likelihood model and the one-sigma spread from a random sample of posterior models is similar in width to the red model line. Neither black hole spin $a$ or inclination $\theta_{\rm{inc}}$ are constrained by the data, but the black hole mass is estimated to be $\log(M_{\rm{AD}}/M_{\odot}) = 10.31^{+0.17}_{-0.14}$.}
    \label{fig:slimbh_SE_fauxtometry}
\end{figure*}

In Figure \ref{fig:slimbh_SE_fauxtometry}, we present a corner plot for the parameter chain from the \texttt{slimbh} model fits to synthetic photometry using the SE virial mass estimate with 0.5 dex width as the prior, which is our default model for future discussions, unless otherwise stated. A black hole mass estimate of $\log(M_{\rm{AD}}/M_{\odot}) = 10.31^{+0.17}_{-0.14}$ is measured from the $M_{\rm{BH}}$ posterior, which has a log-normal shape and a slightly extended high-mass tail. It's clear that neither the black hole spin $a$ nor inclination $\theta_{\rm{inc}}$ are well-constrained by the data, but lower spin and face-on orientation solutions are preferred. Notably, while the slim disc model solutions are clearly skewed towards face-on orientations which we consider intrinsically more probable, the thin disc inclination posteriors are more biased towards edge-on. This further supports that, for extremely luminous QSOs such as J2157--3602, the slim disc model is a more realistic representation of the accretion disc spectra. The convergence between the emergent spectra of the two models suggest that the thin disc model is a good approximation for black holes with low Eddington ratios. To test the sensitivity of our result to the synthetic photometry, we consider variations to the data created by changing the \feii\ template, the \feii\ flux threshold, or the power-law continuum slope applied to the Selsing template, and find no appreciable effect on the resulting black hole mass estimate.

From fitting the photometry alone, we find a 0.05$-$0.10 dex greater $M_{\rm{AD}}$ compared to synthetic photometry, because the additional flux from broad emission-lines boosts the perceived luminosity of accretion disc models. The results from the template-corrected photometry are intermediate between that of the observed photometry and synthetic photometry, with a small but consistent correction of $\Delta M_{\rm{AD}} \sim 0.05$ dex over results from synthetic photometry. However, depending on the properties of an individual QSO with respect to the template, this adjustment may be sufficient to offset the higher mass bias from photometry. The \texttt{kerrbb} thin disc model estimates are also systematically greater than \texttt{slimbh} slim disc models, suggesting that the thin disc assumption may be partially responsible for an overestimation of $M_{\rm{AD}}$ in \citetalias{Campitiello_2020}, although the difference shown for this target is minimal. It remains to be seen whether the discrepancy is greater at lower black hole masses as was the case in \citetalias{Campitiello_2020}.

Between prior probability distributions, the SE virial mass estimate with its 0.5 dex uncertainty is not a strong prior, producing results similar to the uninformative prior. In contrast, the steep high-mass end of the QSO mass function favours $M_{\rm{AD}}$ estimates over 0.10 dex lower. Measurements presented in Table \ref{tab:ADmasses} imply that the rms uncertainty of $M_{\rm{AD}}$ from an unconstrained spin and inclinations below $65^{\circ}$ is less than 0.2 dex. Relaxing the inclination angle limitation or enforcing a more stringent inclination constraint produces black hole mass estimates within 0.6$\sigma$, which implies that our result is not sensitive to the assumed opening angle.

Despite the indications of the spin posterior distribution, one might expect that the spin of the J2157--3602 black hole is fairly high due to its inferred history of prolonged, ordered accretion, in order for it to have grown to its observed mass in roughly 1 Gyr \citep[e.g.][]{Volonteri_2005, Barausse_2012, Walton_2013}. If we used a prior of $a > 0.6$, the $M_{\rm{AD}}$ estimate would be consistent with the unconstrained case, at $\log(M_{\rm{AD}}/M_{\odot}) = 10.39^{+0.19}_{-0.10}$. Independent constraints on degenerate parameters, such as an estimate of spin from the Fe-K$\alpha$ broad lines in the X-ray \citep[e.g.][]{Reynolds_2008, Reynolds_2021}, have the potential to improve the black hole mass estimates from AD fitting. However, as estimated in the later Section \ref{sec:eddrat_radeff}, the radiative efficiency is fairly low, implying lower spins ($a \lesssim 0.5$) in concordance with the MCMC result and that the growth history of J2157--3602 may have involved periods of chaotic accretion, suppressing its spin and radiative efficiency, while maintaining high mass accretion rates \citep{Zubovas_2021}. In this study, we remain agnostic about the spin distribution, preferring to keep the spin prior uniform across the entire positive spin parameter space, which produces a conservative estimate of the precision by which the black hole mass can be measured.

For the results of the thin disc \texttt{kerrbb} model, if the disc power from the torque at the inner boundary of the accretion disc were assumed to be comparable to the disc emission, then the resulting mass estimate using synthetic photometry with the SE prior would be $\log(M_{\rm{AD}}/M_{\odot}) = 10.61^{+0.15}_{-0.21}$, which is 0.18 dex higher than with zero torque. As the \texttt{kerrbb} mass estimates are already higher than the virial and \texttt{slimbh} masses, including the effect of a nonzero torque further exacerbates the disparity. We also consider an alternative slim disc \texttt{slimbh} model with the $\alpha$-viscosity parameter set to 0.1 instead of 0.01, which is a high estimate for AGN accretion discs \citep[e.g.][]{Starling_2004}. The mass estimate with the higher $\alpha$-viscosity is $\log(M_{\rm{AD}}/M_{\odot}) = 10.31^{+0.16}_{-0.15}$, virtually indistinguishable from the original estimate. This is expected as the viscosity parameter has little effect on the emergent spectrum aside from boosting the spectrum to higher energies through electron scattering \citep{Kawaguchi_2003}. However, as discussed previously, Compton effects are unlikely to be significant in the regime of supermassive black holes accreting at sub-Eddington rates. To test this assumption, we utilise our Bayesian MCMC AD fitting method with the slim disc model to constrain the hardening factor, finding $f_{\rm{col}} = 1.00^{+0.25}_{-0.00}$ from the posterior distribution. Therefore, the hardening factor is consistent with unity as we have assumed.

When the limb darkening effect is disabled, we find $\log(M_{\rm{AD}}/M_{\odot}) = 10.33^{+0.15}_{-0.12}$ using the \texttt{slimbh} model with the SE prior. The black hole mass estimate is marginally increased and the mass posterior distribution is slightly narrower. However, the difference between the mass estimates with and without limb darkening depends on the shape of the inclination posterior. For J2157--3602, we find face-on solutions are generally favoured by the slim disc model and the thin disc model favours inclinations closer to edge-on. In this scenario, the \texttt{kerrbb} mass estimates experience a marginal decrease when limb darkening is disabled and the \texttt{slimbh} mass estimates increase slightly.

\setlength{\extrarowheight}{3pt}
\begingroup
\begin{table}
\caption {\label{tab:ADmasses} Black hole masses, $\log{(\rm{M_{BH}}/\rm{M_\odot})}$, measured by the AD fitting method for thin and slim accretion disc models and three different prior probability distributions. The three sets of measurements are distinguished by the dataset to which the AD model is fit. ``SynPhot" refers to the synthetic photometry created from spectroscopic data, ``Template" refers to the photometric data scaled to the continuum model of an average QSO template \citep{Selsing_2016}, and ``Photometry" refers to the photometry-only data. The Galactic extinction correction is applied to all datasets. Our default AD fitting model, marked in bold font and referenced throughout this paper, is the \texttt{slimbh} model fit to synthetic photometry using the Single-Epoch prior.} 
\begin{tabular}{lccc}
\hline \hline
 Model (Data) / Prior & Flat & Single-Epoch & Mass Function \\
 \hline
\texttt{kerrbb} (SynPhot) & $10.44^{+0.11}_{-0.25}$ & $10.43^{+0.12}_{-0.21}$ & $10.29^{+0.17}_{-0.15 }$ \\
\texttt{slimbh} (SynPhot) & $10.32^{+0.17}_{-0.15}$ & $\mathbf{10.31^{+0.17}_{-0.14}}$ & $10.21^{+0.18}_{-0.10}$ \\
\hline
\texttt{kerrbb} (Template) & $10.49^{+0.12}_{-0.24}$ & $10.47^{+0.12}_{-0.22}$ & $10.31^{+0.20}_{-0.13}$ \\
\texttt{slimbh} (Template) & $10.37^{+0.17}_{-0.17}$ & $10.35^{+0.17}_{-0.15}$ & $10.27^{+0.16}_{-0.13}$ \\
\hline
\texttt{kerrbb} (Photometry) & $10.51^{+0.13}_{-0.23}$ & $10.51^{+0.10}_{-0.24}$ & $10.35^{+0.19}_{-0.14}$ \\
\texttt{slimbh} (Photometry) & $10.39 ^{+0.18}_{-0.16}$ & $10.36^{+0.20}_{-0.14}$  & $10.27^{+0.19}_{-0.11}$ \\
\hline \hline
\end{tabular}
\end{table}
\endgroup
\setlength{\extrarowheight}{0pt}

\subsubsection{Dust extinction} \label{sec:dust_ext}
The average QSO with HiBAL features typically exhibits some reddening due to host galaxy dust \citep[e.g.][]{Brotherton_2001, Reichard_2003}. The dust extinction inferred from QSOs at $z < 2.2$ is consistent with the SMC extinction curve \citep{Richards_2003, Hopkins_2004}. At higher redshifts, the extinction curve has a tendency to flatten at $\lambda \leq 2000$\AA, indicating a different dust production mechanism or a different physical mechanism by which dust is processed into the ISM \citep{Gallerani_2010}, although an SMC extinction law is still consistent with the data in most cases \citep[e.g.][]{Krawczyk_2015}. 

In order to quantify the effects of host galaxy dust attenuation, we assume the empirically inferred mean extinction curve from \citet{Gallerani_2010} with an SMC-like R$_{\rm{V}} = 2.7$ \citep{Bouchet_1985} The mean reddening measured for QSOs with HiBAL features is $\Delta$E$_{\rm{B-V}} \sim 0.02$ relative to an ensemble average QSO composite, which is itself reddened by $\Delta$E$_{\rm{B-V}} \sim 0.004$ compared to a non-BAL composite \citep{Reichard_2003}. However, individual measurements show a wide distribution, with a sizable fraction (30--40\%) of HiBALs less reddened than a similar sample of non-BALs \citep{Reichard_2003}. Thus, we do not refer to the mean case to infer host galaxy reddening for an individual target, especially not for a target as extreme as J2157--3602. However, to test how significant dust reddening would affect our measurements, we choose E$_{\rm{B-V}}$ = 0.08, which is fairly substantial dust extinction for a QSO \citep{Krawczyk_2015}. This value was also chosen to keep the disc luminosity sub-Eddington and constrained by the slim disc model.

An AD fit to the de-reddened synthetic photometry results in a black hole mass estimate of $\log(M_{\rm{AD}}/M_{\odot}) = 10.15^{+0.14}_{-0.19}$, which is $\Delta \log(M_{\rm{AD}}/M_{\odot}) = -0.16$ for $\rm{E}_{\rm{B-V}} = 0.08$. Increasing the host galaxy extinction has two competing effects on the AD fit: the true luminosity of the continuum rises and the intrinsic continuum becomes bluer. The former favours higher black hole masses, while the latter implies higher accretion disc temperatures, which favours lower black hole masses. Overall, the influence from a bluer continuum is stronger, resulting in an AD model with a less massive black hole, but higher luminosity and Eddington ratio. 

The SE virial black hole mass estimate with host galaxy extinction would be increased by the boosted 3000\AA\ monochromatic luminosity, which implies a larger BLR radius. Assuming the measured FWHM does not change appreciably, the extinction law implies 
\begin{equation}
    d \log(M_{\rm{SE}}/M_{\odot})/d(\Delta \rm{E}_{\rm{B-V}}) = 0.248 \,\rm{R}_{\rm{V}} \left(A_{\textrm{3000\AA}}/A_{\rm{V}}\right)\,,
   \label{eq:SE_extinction}
\end{equation}
where $\left(A_{\textrm{3000\AA}}/A_{\rm{V}}\right)$ is evaluated from the mean extinction curve and the extinction of the object is measured relative to the mean extinction of the SE calibration sample, meaning $\Delta E_{\rm{B-V}} = E_{\rm{B-V, obj}} - E_{\rm{B-V, cal}}$. Using the \citet{Gallerani_2010} curve with $\left(A_{\textrm{3000\AA}}/A_{\rm{V}}\right) = 1.86$, we find $d \log(M_{\rm{SE}}/M_{\odot})_{\rm{SE}}/d(\Delta \rm{E}_{\rm{B-V}}) = +1.2$. Thus, increasing host galaxy extinction causes our two independent $M_{\rm{BH}}$ measurement methods to diverge, which implies our source is likely to be minimally extinguished. In further discussions, we use the $M_{\rm{BH}}$ estimates without host galaxy extinction, specifically $\log(M_{\rm{AD}}/M_{\odot}) = 10.31^{+0.17}_{-0.14}$ from the default slim AD fitting model.

\subsection{Bolometric Luminosity}
Using the myriad of accretion disc models from the MCMC posterior, we integrate all of the resulting SEDs with an anisotropy correction to derive a bolometric luminosity of $\log{(L_{\rm{bol}}/\rm{erg\,s^{-1}})} = 47.87 \pm 0.10$, implying a bolometric correction factor of $k_{3000\text{\AA}} \sim 1.62$, following $L_{\rm{bol}} = k_{\lambda}L(\lambda)$. This differs from empirically derived bolometric correction factors, such as $k_{3000\text{\AA}} \sim 5.15$ \citep{Richards_2003}, $k_{3000\text{\AA}} \sim 3.33$ \citep{Runnoe_2012}, $k_{3000\text{\AA}} \sim 3.2$ \citep{Trakhtenbrot_2012}, and $k_{3000\text{\AA}} \sim 1.84$ \citep{Netzer_2019}, from which we estimate $\log{(L_{\rm{bol}}/\rm{erg\,s^{-1}})} = 48.17 \pm 0.16$ using all four correction factors. We provide a detailed description of the bolometric luminosity measurement in Appendix \ref{appendix:bol-lum}, and compare it with commonly used empirical bolometric corrections. We also derive and present mean flux anisotropy corrections for samples of QSOs in Appendix \ref{sec:appendix-anisotropy}.

\subsection{Eddington Ratio and Radiative Efficiency} \label{sec:eddrat_radeff}
The Eddington ratio, $\epsilon$, is calculated by the ratio of the bolometric luminosity and the Eddington luminosity for a particular black hole mass, $L_{\rm{bol}} = \epsilon L_{\rm{Edd}}$, where $L_{\rm{Edd}} \approx 1.26 \times 10^{38} (M_{\rm{BH}}/M_{\odot})$ erg s$^{-1}$. From the AD mass estimate, $\log(M_{\rm{AD}}/M_{\odot}) = 10.31^{+0.17}_{-0.14}$, we measure $\epsilon = 0.29^{+0.11}_{-0.10}$. As discussed in \citet{onken_2020_J2157}, the relatively low Eddington ratio suggests that the extreme luminosity of J2157--3602 is a consequence of the size of its black hole. 

Other studies of radiative efficiency for individual QSOs, defined as $\eta$ in $L_{\rm{bol}} = \eta \dot{M}c^2$, are also based on accretion disc models \citep[e.g.][]{Davis_2011, Trakhtenbrot_2014, Trakhtenbrot_2017}. In order to address the degeneracy with inclination, these studies assumed $\cos({\theta_{\rm{inc}}}) = 0.8$, functionally a Dirac $\delta$-function prior in the MCMC context. In the absence of an independent constraint on the inclination, we use the black hole mass prior determined by the SE virial mass estimate. The \texttt{kerrbb} thin disc model with four parameters ($M_{\rm{BH}}$, $\dot{M}$, $a$, $\theta_{\rm{inc}}$) is considered, because the disc luminosity is parameterised in terms of the mass accretion rate. From the fit to synthetic photometry, we determine the MCMC posterior mass accretion rate to be $\dot{M} = 144^{+84}_{-48} \, M_{\odot} \,\rm{yr}^{-1}$, suggesting that the radiative efficiency is $\eta = 0.09^{+0.05}_{-0.03}$, which is consistent with estimates based on the So\l{}tan argument \citep[e.g][]{Yu_2002}.

\subsection{Potential of AD Fitting}
Recently, the potential for AD fitting to complement other black hole mass measurements, such as the SE virial mass method, has been confirmed by several studies \citep{Calderone_2013, Capellupo_2015, Mejia_Restrepo_2018, Campitiello_2020}. Here we find that although neither the black hole spin or observed inclination are constrained by the data, we can obtain approximately log-normal posterior distributions of the black hole mass with widths under 0.2 dex. We are also able to use the model SED to estimate anisotropy-corrected bolometric luminosities within 0.1 dex precision. 

As an independent method of estimating black hole masses, the AD fitting method can be used to disentangle cases for which SE virial mass estimates between different broad lines are in disagreement. For instance, the QSO SDSS J102325.31+514251.0, hereafter J1023$+$5142, is measured to host a black hole of mass $\log(M_{\rm{SE}}/M_{\odot})_{\rm{\mgii}} = 10.52 \pm 0.08$ or $\log(M_{\rm{SE}}/M_{\odot})_{\rm{\hbeta}} = 9.58 \pm 0.14$, as estimated from the \mgii\ or \hbeta\ line, respectively \citep{Zuo_2015}. We collect photometric data by crossmatching J1023$+$5142 with AllWISE, 2MASS, and PanSTARRS DR1. We also use the SFD extinction map with the 14\% recalibration \citep{Schlafly_2011}, finding $E(B-V) = 0.013$, coincidentally matching that of J2157--3602. Assuming no host galaxy extinction and by fitting only photometry, we find $\log(M_{\rm{AD}}/M_{\odot}) = 9.55^{+0.20}_{-0.17}$ using the AD fitting method, which strongly disfavours the \mgii-based estimate. As the measured black hole mass scales inversely to extinction, additional host galaxy reddening corrections will not boost the mass estimate to become consistent with $\log(M_{\rm{SE}}/M_{\odot})_{\rm{\mgii}}$. A more comprehensive description of the fits to J1023$+$5142 photometry can be found in Appendix \ref{sec:appendix-J1023+5142} alongside a discussion of its spectral properties, measured bolometric luminosity, Eddington ratio, and radiative efficiency.

From the \citet{Shen_2011} QSO catalogue of 105,783 QSOs, we find 1599 targets with $\log(M_{\rm{SE}}/M_{\odot}) > 8$ and a difference between the \hbeta\ and \mgii-based virial mass estimate of over 0.5 dex. Applying the same criteria to the \citet{Rakshit_2020} catalogue, we find 1122 such QSOs. In a forthcoming study, we will use AD fitting to investigate targets for which the SE mass estimates are discrepant.

We showed in Section \ref{sec:BHmassAD} that a host galaxy extinction correction has diverging effects on the SE and AD fitting mass estimate. Hence, they can be used together to constrain the magnitude of the host galaxy extinction. A unique advantage of AD fitting is that it can also be used to constrain the radiative efficiency of individual QSOs, as shown in Section \ref{sec:eddrat_radeff}.

Having explored the AD fitting approach extensively on a single target, J2157--3602, we will apply the method to much larger samples in future studies. Samples of luminous QSOs with high-quality spectra, such as XQ-100 \citep{Lopez_2016} and XQR-30 \citep[e.g.][]{Zhu_2021, Lai_2022}, will enable us to produce comparisons between AD fitting and the more established SE virial mass estimate, particularly for supermassive ($\log(M_{\rm{BH}}/M_{\odot}) > 8$) black holes at high redshift.

\section{Conclusions} \label{sec:conclusion}
In this study, the survey photometry and high-quality medium-resolution spectra of the extremely luminous QSO SMSS J215728.21$-$360215.1 enabled us to compare two complementary methods of estimating black hole masses. The main results are as follows:

\begin{itemize}
    \item An updated black hole mass of $\log(M_{\rm{SE}}/M_{\odot})_{\rm{\mgii}} = 10.33 \pm 0.5$ was obtained from the \mgii\ line using multiple empirical and semi-empirical templates to constrain the surrounding \feii\ emission. This estimate is 0.2 dex lower than an earlier analysis for J2157--3602 \citep{onken_2020_J2157}. 
    \item The best performing model from fitting the accretion disc using MCMC suggested a black hole mass of $\log(M_{\rm{AD}}/M_{\odot}) = 10.31^{+0.17}_{-0.14}$, in concordance with the SE estimate. 
    \item We measured the anisotropy-corrected bolometric luminosity to be $\log{(L_{\rm{bol}}/\rm{erg\,s^{-1}})} = 47.87 \pm 0.10$ from the SEDs of the accretion disc models. We also measured the monochromatic luminosity at 3000\AA\ to be $\log{(L(\rm{3000\text{\AA}})/\rm{erg\,s^{-1}})} = 47.66 \pm 0.01$, from which we inferred $\log{(L_{\rm{bol}}/\rm{erg\,s^{-1}})} = 48.17 \pm 0.16$, using bolometric corrections from \citet{Richards_2003}, \citet{Runnoe_2012}, \citet{Trakhtenbrot_2012}, and \citet{Netzer_2019}. We use the bolometric luminosity measured from the accretion disc SED to further derive the Eddington ratio and radiative efficiency. 
    \item The Eddington ratio was calculated to be $\epsilon = 0.29^{+0.11}_{-0.10}$, based on the AD model estimates of the BH mass and bolometric luminosity. 
    \item We also estimated the radiative efficiency to be $\eta = 0.09^{+0.05}_{-0.03}$, measured from the thin disc mass accretion rate posterior distribution. 
\end{itemize}

We demonstrated the utility of fitting accretion disc models in characterising basic properties of high-mass black holes in QSOs. As an independent estimate of black hole mass, this method can be used to complement the SE virial mass estimate, especially at high redshift where other methods are not as easily accessible. The key findings of the AD fitting method are the following:
\begin{itemize}
    \item Both black hole mass $M_{\rm{BH}}$ and the anisotropy-corrected bolometric luminosity $L_{\rm{bol}}$ can be constrained without prior knowledge of the black hole spin or the observed inclination angle. The uncertainties of these measurements are 0.2 dex and 0.1 dex on the mass and luminosity, respectively. Independent constraints on either spin or inclination would improve measurement precision. 
    \item Quantities such as the Eddington ratio and radiative efficiency can be derived from comparing observed properties of the black hole to its intrinsic properties, parameterised as part of the disc models.
    \item It is possible to obtain a reasonable estimate of black hole mass with only broadband photometric data. The offset is found to be 0.1 dex for the target examined in this study, but spectroscopic data can further refine these results.
    \item Host galaxy dust extinction has diverging effects on single-epoch and AD fitting mass estimates. There is potential to exploit this idea to constrain host galaxy extinction for luminous QSOs and the shape of the extinction curve.
\end{itemize}
We plan future studies to investigate a larger sample of luminous QSOs with high-quality spectra, comparing black hole mass estimates between measurements of broad emission-line profiles and the large-scale accretion disc continuum emission.

\section*{Acknowledgements}
We thank the anonymous referee for their constructive comments and suggestions which have improved this manuscript. Additionally, we thank Samuele Campitiello for useful discussions as well as Xiaohui Fan, Feige Wang, and Jinyi Yang for supplying the final reduced spectrum of J2157--3602. We also thank the authors of \citet{Vestergaard_2001}, \citet{Tsuzuki_2006}, \citet{Bruhweiler_Verner_2008}, and \citet{Mejia-Restrepo_2016} for producing and sharing their \feii\ emission templates. 

S.L. is grateful to the Research School of Astronomy \& Astrophysics at Australian National University for funding his Ph.D. studentship.

CAO was supported by the Australian Research Council (ARC) through Discovery Project DP190100252.

This paper is based on observations made with ESO Telescopes at the La Silla Paranal Observatory under programme ID 0104.A-0410(A).

The national facility capability for SkyMapper has been funded through ARC LIEF grant LE130100104 from the Australian Research Council, awarded to the University of Sydney, the Australian National University, Swinburne University of Technology, the University of Queensland, the University of Western Australia, the University of Melbourne, Curtin University of Technology, Monash University and the Australian Astronomical Observatory. SkyMapper is owned and operated by The Australian National University's Research School of Astronomy and Astrophysics. The survey data were processed and provided by the SkyMapper Team at ANU. The SkyMapper node of the All-Sky Virtual Observatory (ASVO) is hosted at the National Computational Infrastructure (NCI). Development and support of the SkyMapper node of the ASVO has been funded in part by Astronomy Australia Limited (AAL) and the Australian Government through the Commonwealth's Education Investment Fund (EIF) and National Collaborative Research Infrastructure Strategy (NCRIS), particularly the National eResearch Collaboration Tools and Resources (NeCTAR) and the Australian National Data Service Projects (ANDS).

This publication makes use of data products from the Wide-field Infrared Survey Explorer, which is a joint project of the University of California, Los Angeles, and the Jet Propulsion Laboratory/California Institute of Technology, and NEOWISE, which is a project of the Jet Propulsion Laboratory/California Institute of Technology. WISE and NEOWISE are funded by the National Aeronautics and Space Administration.

Based on observations obtained as part of the VISTA Hemisphere Survey, ESO Progam, 179.A-2010 (PI: McMahon)

This research uses services or data provided by the Astro Data Lab at NSF’s NOIRLab. NOIRLab is operated by the Association of Universities for Research in Astronomy (AURA), Inc. under a cooperative agreement with the National Science Foundation.

This publication makes use of data products from the Two Micron All Sky Survey, which is a joint project of the University of Massachusetts and the Infrared Processing and Analysis Center/California Institute of Technology, funded by the National Aeronautics and Space Administration and the National Science Foundation.

The Pan-STARRS1 Surveys (PS1) and the PS1 public science archive have been made possible through contributions by the Institute for Astronomy, the University of Hawaii, the Pan-STARRS Project Office, the Max-Planck Society and its participating institutes, the Max Planck Institute for Astronomy, Heidelberg and the Max Planck Institute for Extraterrestrial Physics, Garching, The Johns Hopkins University, Durham University, the University of Edinburgh, the Queen's University Belfast, the Harvard-Smithsonian Center for Astrophysics, the Las Cumbres Observatory Global Telescope Network Incorporated, the National Central University of Taiwan, the Space Telescope Science Institute, the National Aeronautics and Space Administration under Grant No. NNX08AR22G issued through the Planetary Science Division of the NASA Science Mission Directorate, the National Science Foundation Grant No. AST-1238877, the University of Maryland, Eotvos Lorand University (ELTE), the Los Alamos National Laboratory, and the Gordon and Betty Moore Foundation.

Software packages used in this study include \texttt{Numpy} \citep{Numpy_2011}, \texttt{Scipy} \citep{Scipy_2020}, \texttt{Astropy} \citep{Astropy_2013}, \texttt{Specutils} \citep{specutils_2022}, \texttt{Matplotlib} \citep{Matplotlib_2007}, \texttt{emcee} \citep{emcee}, \texttt{corner} \citep{corner} and \texttt{ChainConsumer} \citep{Chainconsumer}.

\section*{Data Availability}
The data underlying this article will be shared on reasonable request to the corresponding author.



\bibliographystyle{mnras}
\bibliography{bibliography} 




\appendix
\section{Bolometric Luminosity}\label{appendix:bol-lum}
Several bolometric corrections have been proposed to transform monochromatic luminosities into QSO bolometric luminosities. At times, a fixed correction is used, $L_{\rm{bol}} = k_{\lambda}L(\lambda)$, where $k_{\lambda}$ is the bolometric correction factor and $L(\lambda)$ represents the monochromatic luminosity, $\lambda L_{\lambda}$ at rest wavelength $\lambda$. From the power-law continuum model of J2157--3602, we measured $\log{(L(\rm{3000\text{\AA}})/\rm{erg\,s^{-1}})} = 47.66 \pm 0.01$.

\subsection{Bolometric corrections} \label{sec:empirical_bolcorr}
Using composite SEDs from 259 SDSS QSOs, \citet{Richards_2006} calculated an average bolometric correction factor of $k_{3000\text{\AA}} = 5.62 \pm 1.14$, although a fixed median value of $k_{3000\text{\AA}} = 5.15$ is more commonly used in practice \citep[e.g.][]{Shen_2011}. However, this value likely overestimates the bolometric luminosity for luminous QSOs, motivating a luminosity-dependent correction which flattens to $k_{3000\text{\AA}} \sim 3.2$ for $L(3000\text{\AA}) > 10^{46}$ erg s$^{-1}$ \citep{Marconi_2004, Trakhtenbrot_2012}, which is derived from QSO spectral templates. 

Another bolometric correction from \citet{Runnoe_2012} is derived from the NIR$-$Xray SED atlas of 63 QSOs \citep{Shang_2011}. Using the best-fitting linear correction with non-zero intercepts and the recommended 25\% luminosity suppression based on the sample's orientation bias \citep[][hereafter \citetalias{Nemmen_2010}]{Nemmen_2010}, the luminosity-dependent bolometric correction for J2157--3602 is then $k_{3000\text{\AA}} \sim 3.33$.

A more recent study based on theoretical calculations of optically thick and geometrically thin accretion discs \citep{Netzer_2019} estimated $k_{3000\text{\AA}} = 25 (L(3000\text{\AA})/10^{42} \rm{erg\,s^{-1}})^{-0.2}$, which is approximately $k_{3000\text{\AA}} \sim 1.84$ for J2157--3602. This study assumed that accretion through the disc is the only energy production mechanism and that the accretion rate is sufficiently sub-Eddington for the disc to remain geometrically thin. Furthermore, the X-ray luminosity from a corona heated to temperatures above that of the AD is assumed to be drawn from the same gravitational energy source and does not affect $L_{\rm{bol}}$. 

The collection of estimated bolometric luminosities with correction factors of (1.84, 3.2, 3.33, 5.15) are $\log{(L_{\rm{bol}}/\rm{erg\,s^{-1}})} =$ (47.93, 48.17, 48.19, 48.38). The mean bolometric luminosity and the standard deviation is then $\log{(L_{\rm{bol}}/\rm{erg\,s^{-1}})} = 48.17 \pm 0.16$. This measurement is well-matched to the previous estimate of $\log{(L_{\rm{bol}}/\rm{erg\,s^{-1}})} \approx 48.2$ \citep{onken_2020_J2157}, which was derived with the \citet{Runnoe_2012} correction. However, the result from independent bolometric corrections spans 0.45 dex and the individual studies assume a mean SED or that a straightforward luminosity dependence is sufficient to extrapolate the full SED shape for every QSO. We adopt an approach in the following section that is independent of the assumed mean SED shape and sample properties used to calibrate the bolometric corrections.

\subsection{Integrated SED} \label{sec:integrate_SED_bol_lum}
The \texttt{kerrbb} thin disc and \texttt{slimbh} slim disc parameterise disc luminosity in terms of the mass accretion rate and Eddington ratio, respectively. However, while the model parameters describe intrinsic properties of the black hole and its accretion disc, the transformation from the model parameterisation to the observed luminosity is not trivial, due to additional effects considered by general relativistic ray-tracing. The mass accretion rate parameter from \texttt{kerrbb} can be converted to luminosity, but the radiative efficiency of the thin disc does not correspond to a standard Keplerian disc \citep{Li_2005}, especially if the torque at the inner boundary is non-zero or limb darkening effects are included. For the slim disc, the Eddington luminosity parameterised in the model does not consider general relativistic effects, and thus only roughly corresponds to the total output disc luminosity \citep{Sadowski_2011}. Therefore, we circumnavigate these potential issues by integrating the model SED to derive the bolometric luminosity. 

Here, we estimate the bolometric luminosity by drawing 1000 random models out of the MCMC posterior distribution from the \texttt{kerrbb} thin disc model fits to synthetic photometry and integrating the SED. We use the thin disc model because the mass accretion rate is a fitted parameter, allowing us to constrain the radiative efficiency and correct for anisotropy. The integrated luminosity, defined by
\begin{equation}
    L_{\rm{iso}} = \int_{0}^{\infty}L_{\nu}d\nu = \int_{-\infty}^{\infty}\ln(10)\nu L_{\nu}d\log(\nu)\,,
   \label{eq:bol_integral}
\end{equation}
is measured to be $\log{(L_{\rm{iso}}/\rm{erg\,s^{-1}})} = 47.99 \pm 0.01$, where $L_{\rm{iso}}$ is the luminosity evaluated under the assumption of isotropy. Although a slim disc is arguably a better model of the accretion disc structure, the difference in the integrated luminosity is only $\log{(L_{\rm{thin}}/L_{\rm{slim}})} = 0.02$. In general, the isotropic-equivalent luminosity, $L_{\rm{iso}}$, is evaluated by 
\begin{equation}
\begin{aligned}
    L_{\rm{iso}} &= \int_{0}^{2\pi} \int_{0}^{\pi} F_{\rm{obs}} d_{L}^2 \sin\theta d\theta d\phi\,,\\
    &= 4\pi d_{L}^2 F_{\rm{obs}}\,,
\end{aligned}
\end{equation}
where $d_{L}^2$ is the luminosity distance and $F_{\rm{obs}}$ is the observed flux density. The isotropic-equivalent luminosity would only be equivalent to the total bolometric luminosity, integrated over all solid angles, if the emission is isotropic or viewed at an intermediate angle where the isotropy assumption is reasonable. If $F_{\rm{obs}}$ is a function of the inclination angle, then we would need to determine a correction for the anisotropy. As the spin and orientation of each thin disc model are known, we can calculate an anisotropy correction to the isotropic luminosity, defined as a ratio of the total bolometric radiative efficiency, $\eta_{\rm{bol}}$, to the isotropic-equivalent radiative efficiency, $\eta_{\rm{iso}}$, such that the total bolometric luminosity, $L_{\rm{bol}}$, is measured by
\begin{equation}
    L_{\rm{bol}} = \frac{\eta_{\rm{bol}}}{\eta_{\rm{iso}}}L_{\rm{iso}} = \dot{M}c^2\int_{0}^{1}\eta(\theta, a) d(\cos\theta)\,,
   \label{eq:iso_efficiency}
\end{equation}
where we construct analytical approximations to estimate $\eta(\theta, a)$ defining,
\begin{equation}
\begin{aligned}
&\eta(\theta, a) = \sum_{i=0}^{\mathcal{N}_{1}} \mathcal{F}_{i}\cos(\theta)^{i} \,,\\
&\mathcal{F}_{i}(a) = \sum_{j=0}^{\mathcal{N}_{2}} \zeta_{j}\log^{j}(1-a)\,,
\label{eq:eta_obs}
\end{aligned}
\end{equation}
and the $\zeta_{j}$ coefficients are presented in Table \ref{tab:coefficients_lbol}, updating the analytical approximation in \citetalias{Campitiello_2018} to include the effects of limb darkening. We caution that our parameterisation differs from that of \citetalias{Campitiello_2018}, but we find this model to produce smoother fits. Coefficients have been provided up to $\mathcal{N}_{1} = 5$ and $\mathcal{N}_{2} = 6$, which are sufficient to estimate $\eta(\theta, a)$ in Equation \ref{eq:eta_obs} to $\sim 0.1\%$ accuracy. 

\begin{figure*}
\includegraphics[width=0.95\textwidth]{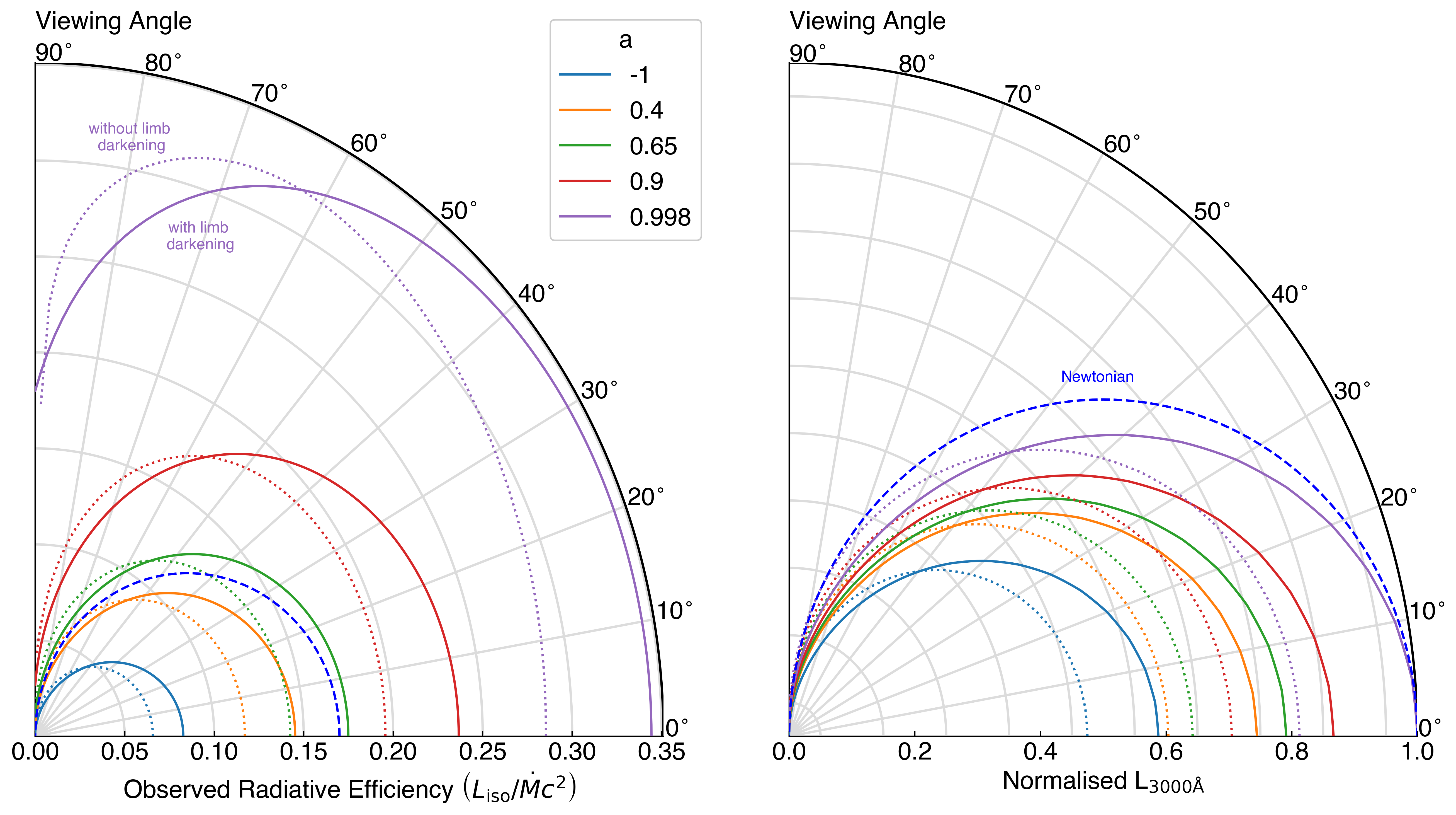}
\caption{Comparison of the observed radiative efficiency (left) and normalised 3000\AA\ monochromatic luminosities (right) at various viewing angles with and without limb darkening, for selected black hole spins from $a=-1$ to $a=1$. The solid (dotted) lines show the result with (without) limb darkening and the Newtonian model, which represents $L \propto \cos\theta$, is shown with the blue dashed lines. On the right panel, the monochromatic luminosities are normalised to the brightest model. With the 3000\AA\ monochromatic luminosity, the shape of the curves remains consistent for the full range of black hole spins.}
\label{fig:ld_comparison}
\end{figure*}

In Figure \ref{fig:ld_comparison}, we compare the observed radiative efficiency and normalised 3000\AA\ monochromatic luminosities with and without limb darkening using the \texttt{kerrbb} thin disc model. With limb darkening, the peak observed luminosity is always at face-on orientations, but without limb darkening, the peak luminosity can be at intermediate viewing angles for close to maximally spinning black holes. Over a large range of viewing angles up to 50--60$^{\circ}$ depending on spin, the models with limb darkening are brighter than similar models without limb darkening. The relationship between the observed radiative efficiency and the viewing angle depends sensitively on spin, but the response of the monochromatic luminosity to orientation is similar at 3000\AA\ regardless of black hole spin. However, the shape of this curve can be quite different at shorter wavelengths, especially beyond the peak of the SED.

Having derived analytical approximations of $\eta(a, \theta)$, we then define the anisotropy correction, $f(\theta, a)$ and a fractional error, $\Delta L/L$, in the isotropic luminosity as,

\begin{equation}
\begin{aligned}
& f(\theta, a) = \frac{\eta_{\rm{bol}}}{\eta_{\rm{iso}}} = \frac{1}{\eta_{\rm{iso}}(\theta, a)}\int_{0}^{1}\eta(\theta', a) d(\cos\theta')\,,\\
&\frac{\Delta L}{L} \equiv \frac{L_{\rm{iso}} - L_{\rm{bol}}}{L_{\rm{bol}}} = \frac{1-f(\theta, a)}{f(\theta, a)}\,,
\label{eq:fractional_error}
\end{aligned}
\end{equation}
where $\Delta L/L$ estimates the error incurred between the observed and total bolometric luminosity of a QSO if the observed emission is assumed to be isotropic. A fractional error of 0 implies no correction and occurs at intermediate viewing angles, such as 60$^{\circ}$ in the Newtonian case or in specific general relativistic scenarios, 56$^{\circ}$ \citep{Netzer_2019} or 66$^{\circ}$ \citepalias{Nemmen_2010} depending on the spin of the system. 

\begin{figure}
	\includegraphics[width=1.0\columnwidth]{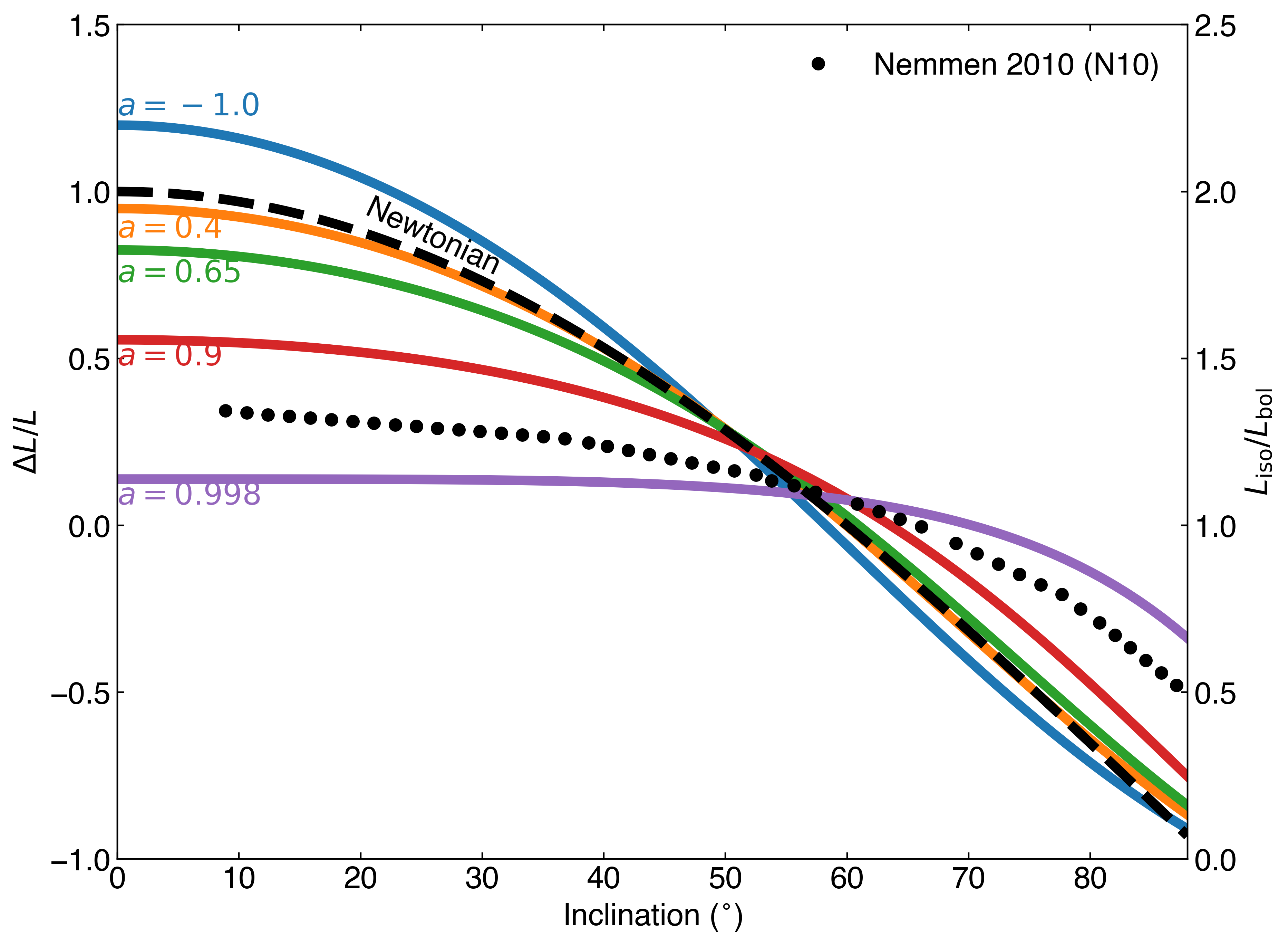}
    \caption{Fractional error incurred between the isotropic luminosity and the total bolometric luminosity of a QSO if the observed emission is assumed to be isotropic. The secondary axis (right) shows the isotropic-equivalent luminosity as a fraction of the bolometric luminosity. Our estimates are derived from the \texttt{kerrbb} thin disc model and contrasted against \citetalias{Nemmen_2010} and the Newtonian case. The mean fractional error is similar to $a=0.4$ if all spins are uniformly probable and to $a=0.65$ if only positive spins are considered.}
    \label{fig:fractional_error}
\end{figure}

In Figure \ref{fig:fractional_error}, we show the fractional error derived from the \texttt{kerrbb} thin disc model across inclination angles for the full range of spins from $a=-1$ to $a=0.998$. We contrast our model with the Newtonian case and the average fractional error from \citetalias{Nemmen_2010}. Much of the parameter space is covered by the highest 10\% of spins. If all spins are uniformly probable, then the $a=0.4$ line best represents the average fractional error. It is also well-approximated by a Newtonian model, except for near face-on and edge-on orientation. If only positive spins are considered, then the $a=0.65$ curve best represents the mean fractional error. These single-spin curves can reproduce the values of their respective mean curves to better than 1\% precision. In Appendix \ref{sec:appendix-anisotropy}, we continue the discussion on the anisotropy correction and derive an average correction for QSO samples with any maximum orientation angle, $\theta_{\rm{max}}$. We create analytical approximations for the correction, $f(\theta_{\rm{max}}, a)$, so that they can be derived for any combination of parameters without evaluating and integrating thin disc model spectra. 

For each of the 1000 models drawn from the MCMC posterior, we generate a unique curve from its black hole spin and evaluate its anisotropy correction for its inclination as in Figure \ref{fig:fractional_error}. After applying each individual correction, we find $\log{(L_{\rm{bol}}/\rm{erg\,s^{-1}})} = 47.87 \pm 0.10$, which implies a bolometric correction factor of $k_{3000\text{\AA}} \sim 1.62$. Methodologically, this procedure is similar to \citet{Netzer_2019}, albeit with the orientation and spin-dependent anisotropy correction instead of a fixed viewing angle assumption of $\sim 56^{\circ}$, and the estimated luminosity is slightly lower as expected due to the shallower mean viewing angle of $\sim 45^{\circ}$. 

Compared to the empirically derived bolometric corrections from Section \ref{sec:empirical_bolcorr}, the integrated AD SED includes only thermal emission from the accretion disc. Only a small fraction of the total radiative output of QSOs is emitted in the X-rays and this fraction also decreases with the UV luminosity of the AGN \citep[the $\alpha_{\rm{OX}}$ correlation; e.g.][]{Elvis_1994, Steffen_2006, Vasudevan_2007, Vasudevan_2009, Liu_2021}. Naturally, we also make the assumption that the thin disc spectra adequately model the observed optical-UV continuum of QSOs. By integrating the locally-blackbody emission, we avoid double-counting reprocessed emission in the infrared or hard X-ray regime. Additionally, we do not depend upon congruence between properties of our target and the mean properties of the samples used to calibrate the bolometric corrections, which can create an order of magnitude difference in the measurement. 

\subsection{Anisotropy Correction} \label{sec:appendix-anisotropy}
In the previous section, we defined an anisotropy correction, $f(\theta, a)$, and its fractional error for any combination of dimensionless black hole spin, $a$, and inclination, $\theta_{\rm{inc}}$. The fractional error estimates the error incurred between the isotropic luminosity and the total bolometric luminosity if the observed emission is assumed to be radiated isotropically. We used the anisotropy correction to estimate the total bolometric luminosity from posterior model SEDs of one QSO with particular spins and observed inclination angles. In this section, we derive the mean anisotropy correction for a sample of QSOs. We first consider the general case of an anisotropy correction that is a function of spin (allowing application to arbitrary spin distributions, e.g., from simulations), then corrections for particular mean spin values. 

Under the orientation classification model for Seyfert galaxies and QSOs \citep{Antonucci_1993}, there's an assumed opening angle of the obscuring torus, where shallower viewing angles enable the observation of the BLR, rendering it a Type 1. Large samples of Type 1 QSOs, where all orientations shallower than the opening angle, $\theta_{\rm{max}}$, are equally likely, will need to consider an average anisotropy correction to transform the observed luminosities into the total bolometric luminosities of the AGN sample. We define this correction, $f(\theta_{\rm{max}}, a)$, as
\begin{equation}
\begin{aligned}
    f(\theta_{\rm{max}}, a) &= \frac{\int_{0}^{1}\eta(\theta, a) d(\cos\theta)}{1-\cos(\theta_{\rm{max}})}\int_{\cos\theta_{\rm{max}}}^{1}\eta(\theta, a)^{-1}d(\cos\theta)\,,
   \label{eq:anisotropy_define}
\end{aligned}
\end{equation}
which is a mean anisotropy correction for a particular black hole spin, $a$, weighted by its probability of being observed at all orientations with $\theta \leq \theta_{\rm{max}}$. As before, the radiative efficiency, $\eta(\theta, a)$, is estimated using Equation \ref{eq:eta_obs}.

We again derive an analytic expression with the following form to estimate the anisotropy correction without evaluating and integrating thin disc model SEDs,
\begin{equation}
\begin{aligned}
&f(\theta_{\rm{max}}, a) = C_{\rm{0}}(\theta_{\rm{max}}) + \sum_{i = 1}^{N_{1}} C_{\rm{i}}(\theta_{\rm{max}}) \log\left(i-a\right) \,,\\
&C_{\rm{i}}(\theta_{\rm{max}}) = \sum_{j = 0}^{N_{2}} D_{\rm{j}} \cos(\theta_{\rm{max}})^{j}\,,
\label{eq:analytic_anisotropy}
\end{aligned}
\end{equation}
where we have provided coefficients, $C_{\rm{i}}$ and $D_{\rm{j}}$, up to $N_1 = 6$ and $N_2 = 8$ in Table \ref{tab:coefficients_anisotropy_analytic}. The provided coefficients are sufficient to estimate $f(\theta_{\rm{max}}, a)$ to within 1\% error for most combinations of dimensionless spin, $a$, and maximum inclination, $\theta_{\rm{max}}$, except for near maximal spin, where the error can be up to 3\%.

\begin{figure*}
\includegraphics[width=0.95\textwidth]{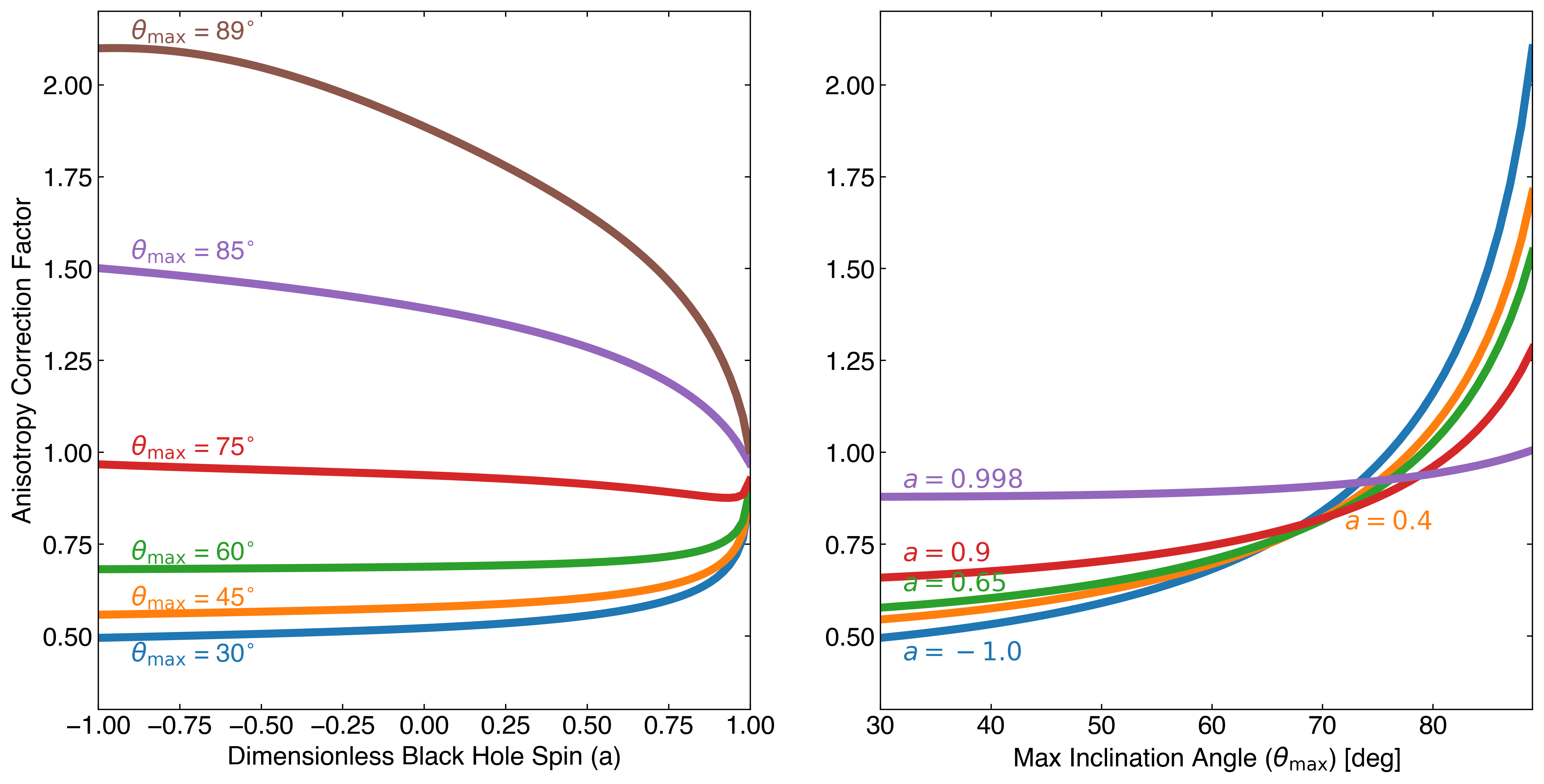}
\caption{Mean anisotropy correction factor for different spins and uniform inclination up to $\theta_{\rm{max}}$ (left) and for maximum inclination angle with uniform spin distribution (right). On the right panel, the $a=0.4$ (or $a=0.65$) spin case best represents the mean curve if all spins (all positive spins) are uniformly probable. For any QSO sample with randomly distributed spins and an estimated opening angle of the obscuring torus, the right panel can be referenced to estimate a mean anisotropy correction.}
\label{fig:anisotropy_corr_figs}
\end{figure*}

In Figure \ref{fig:anisotropy_corr_figs}, we show the mean anisotropy correction for various black hole spins on the left panel and for various maximum inclination angles on the right panel. In the left panel, each of the curves represent a different value of $\theta_{\rm{max}}$. The range of correction factors increases dramatically as $\theta_{\rm{max}}$ approaches edge-on. However, all mean correction factors converge to near $f(\theta_{\rm{max}}, a) \sim 1.0$ for the highest spins. At $60^{\circ} \leq \theta_{\rm{max}} \leq 80^{\circ}$, the mean correction factor is fairly flat with relatively little deviation across all spins, except for near-maximal spin. On the right panel, the mean anisotropy correction is shown for the range of maximum inclination angles. As with Figure \ref{fig:fractional_error}, most of the parameter space is covered by the highest 10\% of spins. 

If all black hole spins in a sample of QSOs are assumed to be uniformly probable, the mean curve is best represented by the $a=0.4$ spin case, and the $a=0.65$ curve best reflects the mean curve for all positive spins. These single-spin curves can reproduce the values of their respective mean curves to better than 1\% precision. For instance, if a sample of QSOs was observed with unknown spins and orientations, but an opening angle of $\theta_{\rm{max}} = 65^{\circ}$ is assumed as in this study, then the right panel of Figure \ref{fig:anisotropy_corr_figs} can be used to evaluate a mean anisotropy correction factor for this sample of $\langle f(65^{\circ}, a)\rangle_{a} = 0.75$, which should then be applied to the isotropic luminosity of QSOs in this sample following Equation \ref{eq:iso_efficiency}. This correction factor is similar to the 25\% bolometric luminosity suppression adopted by \citet{Runnoe_2012} to correct for their orientation bias.

\begingroup
\begin{table*}
\caption {Coefficients of $\zeta_{j}$ used to solve for $\mathcal{F}_{i}(a)$ and applied to Equation \ref{eq:eta_obs} to estimate the observed radiative efficiency $\eta(\theta, a)$ for any combination of observed inclination, $\theta$, and black hole spin, $a$.} \label{tab:coefficients_lbol}
\begin{tabular}{lccccccc}
\hline \hline
  & $\zeta_{0}$ & $\zeta_{1}$ & $\zeta_{2}$ & $\zeta_{3}$ & $\zeta_{4}$ & $\zeta_{5}$ & $\zeta_{6}$  \\ \hline
$\mathcal{F}_{0}$ & 0.0034 & -0.0102 & 0.0121 & 0.0091 & 0.0156 & 0.0057 & 0.0006 \\
$\mathcal{F}_{1}$ & 0.0760 & -0.1092 & 0.0445 & -0.1146 & -0.0167 & 0.0307 & 0.0076 \\
$\mathcal{F}_{2}$ & 0.0698 & -0.0942 & 0.1499 & -0.0735 & -0.6514 & -0.3703 & -0.0578 \\
$\mathcal{F}_{3}$ & -0.0192 & 0.0691 & -0.3491 & 0.9902 & 1.7997 & 0.8101 & 0.1131 \\
$\mathcal{F}_{4}$ & -0.0235 & 0.0603 & 0.1863 & -1.2190 & -1.6659 & -0.6792 & -0.0894 \\
$\mathcal{F}_{5}$ & 0.0106 & -0.0397 & -0.0237 & 0.4420 & 0.5295 & 0.2039 & 0.0258 \\

\hline \hline
\end{tabular}
\end{table*}
\endgroup

\begingroup
\begin{table*}
\caption {\label{tab:Dn_coeff_for_Cn} Coefficients of $D_{\rm{j}}$ used to solve for $C_{\rm{i}}(\theta_{\rm{max}})$ and applied to Equation \ref{eq:analytic_anisotropy} to estimate the anisotropic correction, $f(\theta_{\rm{max}}, a)$, for any sample with black hole spin, $a$, and maximum inclination angle, $\theta_{\rm{max}}$.} \label{tab:coefficients_anisotropy_analytic}
\begin{tabular}{lccccccccc}
\hline \hline
  & $D_{0}$ & $D_{1}$ & $D_{2}$ & $D_{3}$ & $D_{4}$ & $D_{5}$ & $D_{6}$ & $D_{7}$ & $D_{8}$  \\ \hline
$C_{0}$ & 825839.11 & -825654.54 & -12727.77 & 133489.26 & -618406.28 & 1542748.35 & -2134722.89 & 1540498.01 & -452008.97 \\
$C_{1}$ & -145.85 & 145.92 & -0.92 & 3.70 & -11.32 & 22.76 & -27.71 & 18.39 & -5.09 \\
$C_{2}$ & 19815.54 & -19601.21 & -942.79 & -2700.28 & 27793.58 & -86347.80 & 132556.84 & -101623.32 & 31026.11 \\
$C_{3}$ & 852776.56 & -855518.06 & -15310.58 & 350278.39 & -1856395.08 & 4895185.38 & -6978921.44 & 5130863.11 & -1524798.56 \\
$C_{4}$ & 19495660.09 & -19487281.21 & 176344.12 & -2554055.17 & 12705285.15 & -32688850.21 & 46008156.13 & -33560153.75 & 9920635.72 \\
$C_{5}$ & 6247439.48 & -6256063.92 & -400372.22 & 5131351.82 & -24930987.93 & 63518064.50 & -88927307.96 & 64653620.82 & -19069014.39 \\
$C_{6}$ & 8396532.21 & -8393904.69 & 249288.84 & -3018336.82 & 14486023.68 & -36713644.08 & 51252728.94 & -37195437.91 & 10956809.26 \\
\hline \hline
\end{tabular}
\end{table*}
\endgroup

\section{QSO J102325.31$+$514251.0} \label{sec:appendix-J1023+5142}
The QSO J102325.31$+$514251.0, $z = 3.477$, was measured to host a black hole of mass $\log(M_{\rm{SE}}/M_{\odot})_{\rm{\mgii}} = 10.52 \pm 0.08$ using the SE virial mass estimate with the \mgii\ line \citep{Zuo_2015}, which would make it one of the most massive black holes ever measured. However, that estimate is in contention with the \hbeta-based virial mass estimate in the same study which suggested a black hole mass of $\log(M_{\rm{SE}}/M_{\odot})_{\rm{\hbeta}} = 9.58 \pm 0.14$ \citep{Zuo_2015}. These cases where the virial mass estimates from different emission lines differ by more than 0.5 dex are rare, constituting about 3\% of QSOs in the \citet{Shen_2011} QSO catalogue and about 1\% in the \citet{Rakshit_2020} catalogue. 

As an independent method of estimating black hole masses, AD fitting is useful for disentangling these ambiguous cases. Even if only limited to cases for which $\log(M_{\rm{SE}}/M_{\odot}) > 8$, where the $f_{\rm{col}} = 1$ assumption is reasonable (see Section \ref{sec:ADfitting}), we find $> 1000$ QSOs in each of the \citet{Shen_2011} and \citet{Rakshit_2020} QSO catalogues. Here, we show how AD fitting can be used to measure the black hole mass of J1023$+$5142, using only publicly available photometric data. 

\begin{figure*}
	\includegraphics[width=0.9\textwidth]{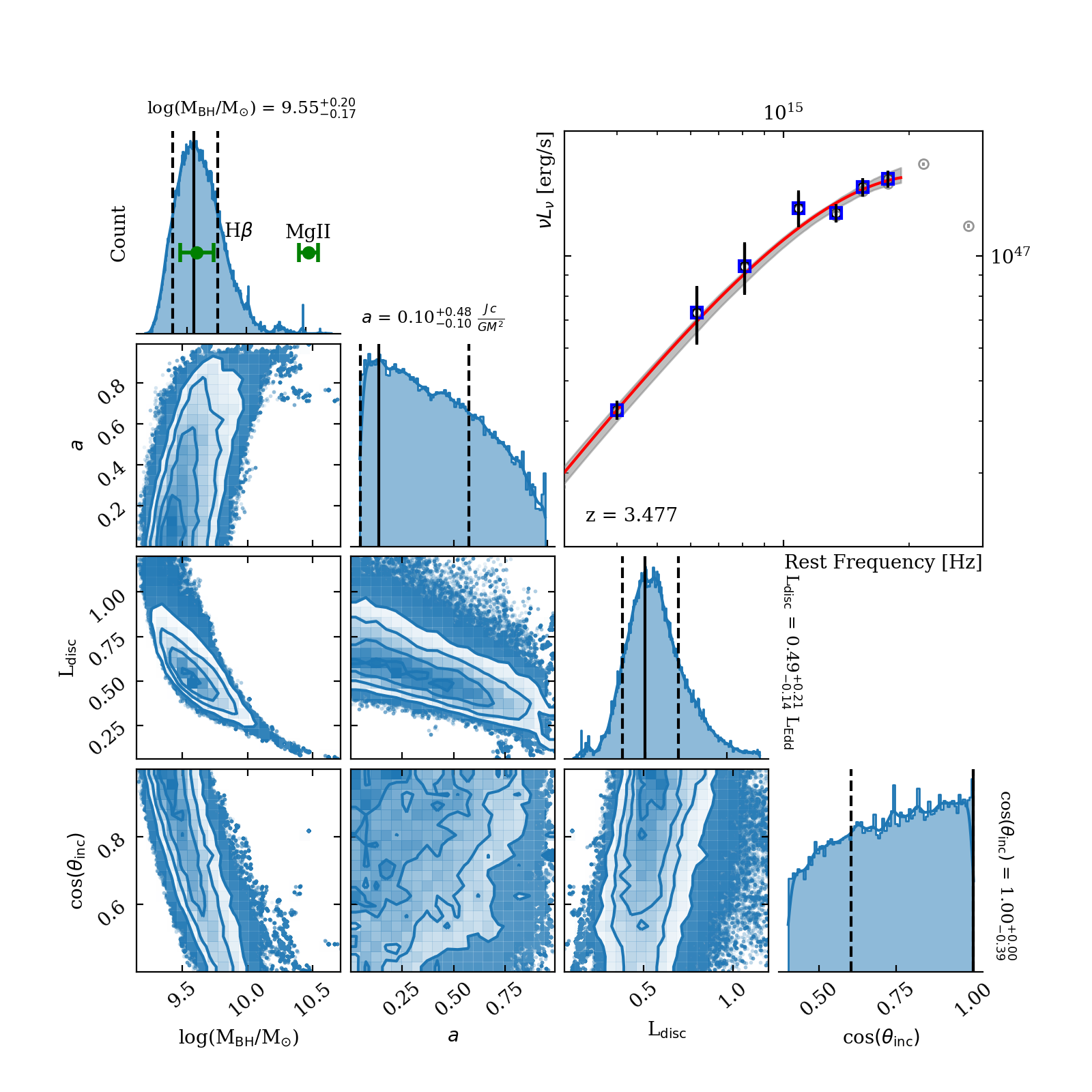}
    \caption{Corner plot for the accretion disc fit to QSO J102325.31+514251.0 photometry using the \texttt{kerrbb} model. The frequency is presented in the rest-frame. The final estimate of each parameter is determined by the maximum likelihood of the posterior distribution, with uncertainties determined by the 68\% iso-likelihood line. The top-right panel shows the photometry in black, and masked data in light grey. The red model is the highest likelihood model and the one-sigma spread from a random sample of posterior models is similar in width to the red model line. The black hole mass is estimated to be $\log(M_{\rm{AD}}/M_{\odot}) = 9.55^{+0.20}_{-0.17}$ and we show each of the SE virial mass estimates in green.}
    \label{fig:J102325.31+514251.0}
\end{figure*}

We crossmatch J1023$+$5142 with AllWISE, 2MASS, and PanSTARRS DR1, obtaining data from the optical to infrared. Following the procedure from Section \ref{sec:ADfitting}, we fit data between rest-frame 1 $\mu$m to 1600\AA. Like J2157--3602, we determine the Galactic extinction based on the rescaled SFD extinction map to be $E(B-V) = 0.013$ and we assume no host galaxy extinction. We adopt the \texttt{slimbh} slim accretion disc models and flat uninformative prior probability distributions for all free parameters: black hole mass $M_{\rm{BH}}$, disc luminosity $L_{\rm{disc}}$, spin $a$, and inclination $\theta_{\rm{inc}}$. Figure \ref{fig:J102325.31+514251.0} presents the corner plot of the MCMC posterior distributions. Although neither spin nor inclination are constrained by the data, the black hole mass estimate is $\log(M_{\rm{AD}}/M_{\odot}) = 9.55^{+0.20}_{-0.17}$, using the maximum likelihood and uncertainties determined by the 68\% iso-likelihood line. This result strongly disfavors the \mgii-based virial mass estimate, suggesting nearly 1 dex lower mass. If the \hbeta\ mass estimate is unavailable and the \mgii\ SE mass estimate is used as the mass prior, the AD fitting mass estimate is $\log(M_{\rm{AD}}/M_{\odot}) = 9.71_{-0.20}^{+0.19}$, which is sufficient to exclude the \mgii\ result at $> 4\sigma$ even if there is no alternative SE $M_{\rm{BH}}$ estimate.

We calculate the bolometric luminosity of J1023$+$5142 following the methods outlined in Appendix \ref{appendix:bol-lum}, finding $\log{(L_{\rm{bol}}/\rm{erg\,s^{-1}})} = 47.47 \pm 0.17$, 0.3 dex lower than in \citet{Zuo_2015}, which used a fixed correction of $k_{5100\text{\AA}} = 9.26$ \citep{Richards_2006}. Instead, we find our bolometric luminosity to be consistent with $k_{5100\text{\AA}} = 4.68$ and $k_{3000\text{\AA}} = 4.27$. Higher mass BHs such as J1023$+$5142 are more likely to have lower bolometric correction factors \citep[e.g.][]{Trakhtenbrot_2012} than those derived from mean SEDs of lower-mass black holes. With the newly measured black hole mass bolometric luminosity using AD fitting, we find an Eddington ratio of $\epsilon = 0.7^{+0.4}_{-0.4}$. We also estimate the mass accretion rate from the \texttt{kerrbb} thin disc model following Section \ref{sec:eddrat_radeff}, finding $\dot{M} = 52^{+42}_{-23} \, M_{\odot} \,\rm{yr}^{-1}$, which implies a radiative efficiency of $\eta = 0.10^{+0.08}_{-0.05}$, consistent with the So\l{}tan argument \citep[e.g][]{Yu_2002}.

Similar to J2157--3602, the SDSS spectra of J1023$+$5142 shows characteristics of a HiBAL QSO, which are, on average, found to be only slightly reddened by host galaxy dust compared to a non-BAL sample \citep[e.g.][]{Brotherton_2001, Reichard_2003}. We find that, even with photometry alone, increasing the assumed host galaxy extinction decreases the estimated black hole mass from the AD fitting method, so $M_{\rm{AD}}$ could not be boosted to the $\log(M_{\rm{SE}}/M_{\odot})_{\rm{\mgii}}$ mass through an intrinsic reddening correction. Because our $M_{\rm{AD}}$ estimate is not overestimated compared to the H$\beta$ SE virial mass, we find that our assumption of no internal reddening for luminous QSOs is not contradicted.

This result motivates a future study to investigate cases for which the virial mass estimates are discrepant in additional detail using AD fitting. It also demonstrates that AD fitting can be used to discover cases for which the SE virial mass estimate may have resulted in erroneous measurements.

\bsp	
\label{lastpage}
\end{document}